\documentclass[aps,prx,twocolumn,groupedaddress,superscriptaddress,nofootinbib,notitlepage,floatfix]{revtex4-2}
\usepackage{times,bm,bbm,amssymb,amsmath,amsfonts,graphics,graphicx,color,hyperref}
\usepackage[english]{babel}
\usepackage{tabularx}
\usepackage[final]{changes}
\usepackage{svg}
\usepackage{mathtools}

\hypersetup{colorlinks,linkcolor={blue},citecolor={blue},urlcolor= {blue}}
\usepackage{graphicx}
\usepackage{float}
\usepackage{siunitx}
\usepackage{amsmath}
\usepackage{qcircuit}
\hypersetup{colorlinks,linkcolor={blue},citecolor={blue},urlcolor= {blue}}
\newcommand{\ignore}[1]{}
\DeclareFontFamily{OT1}{pzc}{}
\DeclareFontShape{OT1}{pzc}{m}{it}
              {<-> s * [1.25] pzcmi7t}{}
\DeclareMathAlphabet{\mathpzc}{OT1}{pzc}
                                 {m}{it}

\DeclareMathOperator*{\argmin}{argmin}

\newcommand{\LU}{\affiliation{Department of Computer Science, Faculty of Science, Technology and Medicine, University of
Luxembourg, Luxembourg}}

\newcommand{\KAIST}{\affiliation{Department of Physics, KAIST,  Daejeon,  Republic of Korea}}
\newcommand{\IISC}{\affiliation{Department of Instrumentation and Applied Physics,
Indian Institute of Science, 
Bengaluru, India}}

\newcommand{\UHeid}{\affiliation{Institute of Computer Engineering (ZITI), Heidelberg University,
69120 Heidelberg, Germany}}               
\usepackage[T1]{fontenc}
\usepackage{newtxtext,newtxmath}
\begin{document}
\author{Aravind P. Babu}
\email{aravindp@iisc.ac.in}
\LU
\IISC
\author{Seongjin Ahn}
\email{seongjin.ahn@kaist.ac.kr}
\KAIST

\author{Jing Sun}
\UHeid

\author{Augustas G. Landsbergis}
\KAIST
\author{Andrey S. Moskalenko}
\email{moskalenko@kaist.ac.kr}
\KAIST

\author{Marko J. Ran\v{c}i\'{c}}
\email{marko.rancic@uni.lu}
\LU
\UHeid

\title{Simulating the Dicke Model on Qubit-Based and hybrid Qubit-Boson-Based 
Quantum Computers} 
\date{\today}
\begin{abstract}
The Dicke model provides a fundamental description of collective light-matter interactions and has long served as a testbed for exploring a wide range of physical phenomena in quantum optics and condensed matter physics. In this work, we develop a variational framework for investigating the finite-size Dicke model on both fully qubit-based (digital) and hybrid qubit boson based (digital-analogue) quantum computing platforms. We show that the resulting model reproduces the characteristic critical behavior of the Dicke model in the appropriate large-spin limit while remaining suitable for implementation on both classical emulators of quantum computers and actual trapped ion quantum computers, albeit in the case of latter somewhat limited by noise. Finally, we introduce a complementary hybrid qubit-bosonic variational ansatz that directly exploits the bosonic degree of freedom to reduce quantum resources and discuss its potential implementation on hybrid quantum hardware. Our results establish a scalable, symmetry-aware framework for variational quantum simulations of collective light-matter systems and provide a pathway toward efficient simulations of more general spin-boson models on near-term quantum devices.

\end{abstract}
\maketitle
\section{Introduction}
The Dicke model Hamiltonian describes the collective interaction between a set of two-level systems and a single-mode bosonic field \cite{dicke1954}. It has long served as a fundamental framework in quantum optics and condensed matter physics, capturing phenomena such as superradiance \cite{bonifacio1971,rehler1971} and quantum phase transitions \cite{hepp1973,wang1973,carmichael1973}. Building on the theoretical foundations laid by the Rabi model \cite{rabi1937} and subsequent generalizations \cite{forn2019}, numerous theoretical studies have explored the Dicke model using analytical and numerical methods, and several experimental works have realized its key features in platforms such as cold atoms in optical cavities \cite{klinder2015}, trapped ions \cite{safavi2018}, and superconducting qubits \cite{mezzacapo2014digital,lamata2017digital}. Mean-field treatments are widely employed \cite{hepp1973,wang1973} to capture the collective behavior, providing analytical insight into phenomena including the superradiant phase transition in the thermodynamic limit.
Using exchange symmetry of the Dicke model,
it is also possible to compute the ground state numerically 
for a finite $N$ up to few thousands \cite{chen2008}.

However, in experiments, the exchange symmetry can break down, 
because the spin-boson coupling differs slightly from one atom to another.
This requires the computation of the ground state to include the whole Hilbert space, which scales exponentially with the number of two-level atoms and cannot be efficiently simulated even on the most powerful supercomputers (see Appendix \ref{sec:classical_simulability}).
In Ref.~\cite{PhysRevA.109.013715}, such disordered Dicke model has been studied classically with the number of particles around $N = 6 \sim 8$.

Quantum computing has emerged as a promising approach for simulating complex many-body physics problems efficiently. Various works have demonstrated successful simulations of models such as the Bose–Hubbard \cite{jaksch1998,greiner2002}, Ising \cite{friedenauer2008,kim2010,salathe2015}, and Heisenberg \cite{ma2011,salathe2015} Hamiltonians on different quantum computing platforms, with potential applications in materials science, quantum chemistry, and high-energy physics. A range of algorithms suitable for both gate-based quantum computers and analogue quantum simulators have been employed for this purpose. Collective light–matter interactions, in particular, have been explored with quantum simulators in several previous studies, primarily using analogue quantum simulators~\cite{langford2017experimentally,cai2021observation,marquez2024quantum,pedernales2015quantum}. Recent works have further extended these investigations to digital-analog quantum simulations of the Rabi and Dicke models, demonstrating efficient simulation across all coupling regimes and enabling scalable implementations with superconducting circuits~\cite{lamata2017digital,mezzacapo2014digital}.

Variational quantum algorithms (VQAs) have demonstrated considerable potential for addressing many-body physics problems on near-term quantum devices \cite{Yuan2019,Cerezo2021,Bharti2022}. The central idea is to use parameterized quantum circuits to prepare trial states, whose parameters are optimized through a classical feedback loop to minimize a cost function, typically the system’s energy. Among these methods, the variational quantum eigensolver (VQE) has become the most widely studied \cite{Cerezo2021,Bharti2022,Peruzzo2014,McClean2016,Tilly2022,Kandala2017,Tang2021,Shkolnikov2023}. VQE can approximate both ground- and excited-state properties of complex Hamiltonians while requiring less resources (in terms of quantum gates and number of physical qubits) compared to fault-tolerant approaches such as the quantum phase estimation, making it particularly well suited to noisy intermediate-scale devices. As a complementary method to the analogue quantum simulation, this approach forgoes the requirement of realizing the target Hamiltonian directly in the laboratory, thus enabling the study of a wide variety of previously intractable models \cite{Kokail2019}. Originally developed for quantum chemistry applications \cite{Peruzzo2014,OMalley2016}, where the exponential growth of the Hilbert space renders accurate molecular simulations infeasible for classical computers, VQE has been also used to study a variety of condensed-matter problems, including spin models \cite{Kandala2017}, Hubbard-type systems \cite{Reiner2019,Cade2020}, and frustrated magnets \cite{Liu2019,Kattemolle2022}. Recently, quantum phase transition of a local many-body Hamiltonian has been studied using VQE, targeting the transverse-field Ising model with boundary fields~\cite{variational-algo-q-phase-trans-ising-2025}, demonstrating the gradual build up of the order parameter toward discontinuity at the critical value of the boundary field. 

Variational approaches have also been extended to light-matter systems; in particular, in Ref.~\cite{VQE_ultrastrong_lightmatter_PhysRevResearch} a polaron-inspired variational ansatz has been introduced for a multi-mode Dicke model and demonstrated proof-of-principle ground-state preparation on superconducting quantum processor but for few qubits. However, in contrast to our work, the manuscript in question only discusses the ultra-strong coupling limit. Somewhat related work \cite{PhysVQE_cQED} has also explored digital quantum simulations of mixed spin--boson Hamiltonians for cavity-QED systems, or to be more specific, the Jaynes–Cummings model in two coupled cavities.

In this work, we develop a variational quantum framework for studying the finite-size Dicke model on both fully digital and hybrid quantum computing platforms. To enable implementation on the digital quantum processors, we reformulate the Dicke Hamiltonian through an inverse Holstein-Primakoff transformation \cite{Holstein1940,Matsubara1956,Wilson2016}, which encodes the bosonic cavity mode into a finite number of qubits. Approximation of  the resulting representation yields an effective finite-dimensional Hamiltonian, referred to throughout this work as the spin-Dicke model. Based on this representation, we construct symmetry-preserving variational ansätze for the variational quantum eigensolver by explicitly exploiting the parity, time-reversal, and exchange symmetries of the finite-size Hamiltonian. We first assess their performance through full VQE simulations on an ideal statevector simulator, demonstrating accurate ground- and excited-state calculations over a broad range of coupling strengths. We further perform a systematic study of the proposed ansätze by examining their analytical structure and the behavior of the optimized variational parameters across several representative cases. This analysis provides physical insight into how the symmetries of the Hamiltonian shape the variational optimization, identifies redundant variational parameters that can be eliminated, and establishes general design principles for constructing scalable ansätze.

To examine the experimental viability of the proposed approach, we prepare the classically optimized variational states on a trapped-ion quantum processor, whose all-to-all qubit connectivity naturally matches the interaction topology required for the Dicke model. Finally, we introduce a complementary hybrid qubit--bosonic variational ansatz, in which the atomic degrees of freedom are encoded in qubits while the cavity mode is treated directly as a bosonic mode. By avoiding a qubit encoding of the cavity, this approach reduces the required number of qubits and circuit depth.

\section{Dicke Model}
The Dicke model describes 
an ensemble of $N$ two-level atoms that collectively interact with a single cavity mode. The corresponding Hamiltonian is given by
\begin{align} 
\label{eq:hamiltonian_homo}
    H = \omega a^{\dagger}a + \sum_{n=1}^{N} \omega_{0}\frac{\sigma_{an}^{z}}{2} + \sum_{n=1}^{N}\frac{\lambda}{\sqrt{N}} (a^{\dagger}+a)\sigma_{an}^{x}, 
\end{align}
where $a^\dagger$ and $a$ are the bosonic creation and annihilation operators of the cavity mode with frequency $\omega$, and $\sigma_{an}^{\alpha}$ ($q = x, y, z$) is a Pauli operator acting on the $n$th atom.  The parameter $\omega_{0}$ denotes the energy splitting of the two-level atoms, and $\lambda$ represents the atom-photon coupling strength per atom \cite{Blaha2022}. We choose $\hbar =1$. The standard Dicke model in Eq.~(\ref{eq:hamiltonian_homo}) assumes a homogeneous setting, where all atoms have the same frequency $\omega$ and are coupled to the cavity mode with identical strength $\lambda$. In realistic physical systems, however, achieving perfect homogeneity is challenging: in practice, one often encounters inhomogeneous broadening of atomic frequencies or spatially varying coupling strengths. To capture these effects, the model can be generalized to the inhomogeneous Dicke Hamiltonian by allowing site-dependent frequencies and couplings
\begin{align} 
\label{eq:hamiltonian_inhomo}
   H = \omega a^{\dagger}a+ \sum_{n=1}^{N} \omega_{n}\frac{\sigma_{an}^{z}}{2} + \sum_{n=1}^{N}\frac{\lambda_{n}}{\sqrt{N}} (a^{\dagger}+a)\sigma_{an}^{x}, 
\end{align}
 where $\omega_{n}$ for $1 \le n \le N$ denotes the transition frequency of the $n$th atom, and $\lambda_{n}$ represents its individual coupling strength to the cavity field. The Dicke Hamiltonian has the parity symmetry, which corresponds to invariance under the joint transformation
\begin{equation}
\mathcal{P}:  a \rightarrow -a, \quad \sigma_{an}^{x} \rightarrow -\sigma_{an}^{x}.
\end{equation}
It is associated with a conserved quantity, the parity of the total number of excitations,  $\Pi = \exp\left[ i \pi \left( a^\dagger a + \sum_{n=1}^{N} \frac{\sigma_{an}^{z} + 1}{2} \right) \right]$. Note that in this case, the total number of excitations is not conserved, since the Hamiltonian includes counter-rotating terms. However, the total number of excitations always changes in multiples of two, thereby preserving the parity symmetry. 

When varying the coupling strength $\lambda$, the Dicke model exhibits a second-order phase transition in the thermodynamic limit $N \to \infty$, known as the superradiant phase transition, which occurs when the collective coupling exceeds a critical value \cite{hepp1973,wang1973,emary2003}. In this regime, the system transitions from the normal phase, where the bosonic field has no macroscopic occupation, to the superradiant phase, characterized by spontaneous symmetry breaking.

\begin{figure}
    \centering
    \includegraphics[width=\linewidth]{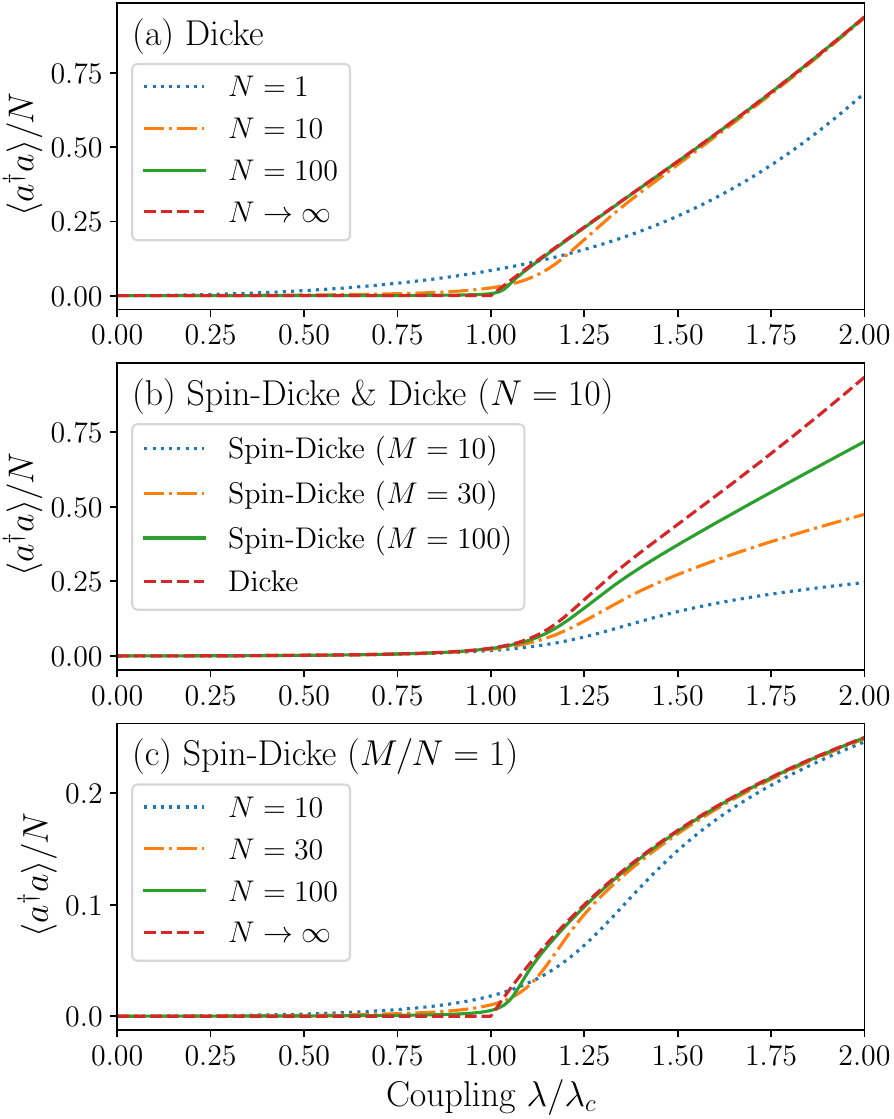}
    \caption{
        Average 
    number of cavity excitations per atom, 
    $\langle a^{\dagger}a\rangle/N$, in dependence on the normalized coupling strength $\lambda/\lambda_c$ [$\lambda_c$ is introduced after Eq.~\eqref{eq:avg-adag-a-over-N-sd-thermlim}]  in the ground state of
        (a) the Dicke model, (b) the spin-Dicke model 
        for $N=10$ and several values of $M$ 
        compared to the Dicke model with $N=10$, and (c) the spin-Dicke model for several $N$ with fixed $M/N=1$.
        For the spin-Dicke model, the average boson number is computed via the connection, $\langle a^{\dagger}a\rangle \mapsto \langle S_{z}\rangle + M/2$, given in Eq.~(\ref{eq:inv-hp-adag-a}) with $s = M/2$.
    }
    \label{fig:avg-boson-num-dicke-and-spin-dicke}
\end{figure}

For a small $N$, the system undergoes a smooth transition. The phase boundary becomes well-defined in the thermodynamic limit \cite{Fisher1972,Cardy1988,Binder1992}. However, even for a finite $N$ we can still observe precursors of the normal-to-superradiant phase transition: sharp crossovers in observables such as the average photon number, which become increasingly pronounced as $N$ increases \cite{emary2003,Vidal2006,chen2008}. These finite-size effects are particularly relevant for near-term quantum simulators, where only small or intermediate numbers of qubits can be realized. In Fig.~\ref{fig:avg-boson-num-dicke-and-spin-dicke}(a), we illustrate the gradual approach to the thermodynamic limit of the average boson number in the ground state of the Dicke model with increasing $N$. In this work, we mainly consider the finite size cases which can be simulated on a quantum computer. 

\section{Spin-Dicke model}
Solving the Dicke Hamiltonian with the variational quantum eigensolver (VQE) requires expressing both the bosonic and the spin operators in the qubit basis.  
In particular, the bosonic Hilbert space should be truncated to have a finite dimension.
After truncation, bosonic operators can be mapped to qubit operators in several ways, each with different resource requirements and circuit complexities~\cite{PhysVQE_cQED, somma2003quantumdirectencoding, huang2021qubitization}. In this work, we use the inverse Holstein-Primakoff transformation (HPT) to map the bosonic mode to the qubit. 


The inverse HPT expresses the bosonic operator in terms of spin-$s$ operators as
\begin{equation}\label{eq:inv-hp-a}
    a \mapsto \frac{1}{\sqrt{s - S_{z}}} S_{-},
\end{equation}
for which the bosonic number operator maps to
\begin{equation}\label{eq:inv-hp-adag-a}
    a^\dagger a \mapsto s + S_{z}.
\end{equation}
Introducing eigenvectors as ${S_{z}|{m}\rangle_{S} = m|{m}\rangle_{S}}$ for ${-s \le m \le s}$,
Eq.~(\ref{eq:inv-hp-a}) results from the observation on the annihilation and the spin ladder operators, $a$ and $S_{-}$, respectively, for which

\begin{equation*}
\begin{split}
    a|{n+1}\rangle_{F}
    &= \sqrt{n+1}|n\rangle_{F},
    \\
    S_{-}|{-s+n+1}\rangle_{S}
    &= \sqrt{n+1}\sqrt{s-(-s+n)}|{-s+n}\rangle_{S}
\end{split}
\end{equation*}
for $0 \le n < 2s$. Here, we see that $\sqrt{n+1}$ is a common factor.
Noting that $\sqrt{s-m}|{m}\rangle_{S} = \sqrt{s-S_{z}}|{m}\rangle_{S}$ since $|{m}\rangle_{S}$ is the eigenvector of $S_{z}$ with eigenvalue $m$, we identify
\begin{equation*}
    \sqrt{s-(-s+n)}|{-s+n}\rangle_{S} = \sqrt{s-S_{z}}|{-s+n}\rangle_{S}.
\end{equation*}
Multiplying by the inverse of $\sqrt{s-S_{z}}$ from the left 
and mapping each Fock state $|n\rangle_{F}$ to the spin state $|{-s+n}\rangle_{S}$, we obtain $a|{n+1}\rangle_{F} \mapsto (s-S_{z})^{-1/2}S_{-}|{-s+n+1}\rangle_{S}$ for $0 \le n < 2s$.
We refer to Appendix~\ref{sec:HPT} for a brief derivation of the transformation.

Note that $2s \equiv M$ is the maximum number of bosons allowed in this truncated representation of the bosonic Hilbert space. In order to well approximate the Dicke eigenstate $|E\rangle$ in the finite basis, $M$ needs to be sufficiently large such that 
    $\mathrm{Prob}[{a^{\dagger}a > M}] = \sum_{n>M} \left| \langle E| n\rangle_{F}\right|^2$
is small. For the ground state and the first few excited states, we expect that such $M$ exists for finite $N$ and coupling strength.
Combined with Eq.~(\ref{eq:inv-hp-adag-a}), this implies that $s+S_{z}$ would take values much lower than $2s$ with high probability for the low-lying eigenstates.
Based on this argument, we rewrite the square root in Eq.~(\ref{eq:inv-hp-a}) as $\sqrt{s-S_{z}} = \sqrt{2s-(s+S_{z})}$ and factor out $\sqrt{2s}$ to expand
\begin{equation}\label{eq:inv-hp-a-expanded}
    a
    \mapsto
    \left[
    1
    +
    \frac{1}{2}
    \frac{s + S_{z}}{2s}
    + \cdots
    \right]
    \frac{S_{-}}{\sqrt{2s}}.
\end{equation}
Since the spin-$s$ operators can be decomposed into $M=2s$ spin-$1/2$ Pauli operators as
\begin{equation*}
\begin{split}
    S_{z} &= \frac{\sigma_{c1}^{z}}{2} + \cdots +\frac{\sigma_{cM}^{z}}{2},\\
    S_{-} &= \sigma_{c1}^{-}+\cdots+\sigma_{cM}^{-},
\end{split}
\end{equation*}
where $\sigma_{cm}^{z} = |{\uparrow}\rangle_{cm}\langle{\uparrow}| - |{\downarrow}\rangle_{cm}\langle{\downarrow}|$ and $\sigma_{cm}^{-} = |{\downarrow}\rangle_{cm}\langle{\uparrow}|$,
the first-order term of the expansion in Eq.~(\ref{eq:inv-hp-a-expanded}) is linear with respect to the Pauli operators, whereas the $k$th-order term contains products of $k$ Pauli operators.
Thus, the spin-boson interaction terms, $(a+a^{\dagger})\sigma_{n}^{x}$, of the Dicke model is $(k+1)$-local if we truncate the expansion up to the $k$th order.
In this work, we take into account the leading-order ($k=1$) term,
\begin{equation}\label{eq:a-to-Sminus}
    a \mapsto \frac{S_{-}}{\sqrt{2s}},
\end{equation}
in the expansion given by Eq.~(\ref{eq:inv-hp-a-expanded}) and control the next-order terms by using a sufficiently large $M = 2s$ such that
\begin{equation*}
    \mathrm{Prob}\left[\frac{s+S_{z}}{2s} < \epsilon\right] > 1 - \delta
\end{equation*}
for given $\epsilon, \delta \ll 1$, where the probability is taken with respect to a low-lying eigenstate of interest.
Applying Eq.~(\ref{eq:a-to-Sminus}) with $2s = M$ to the Dicke Hamiltonian given in Eq.~(\ref{eq:hamiltonian_homo}),
we obtain
\begin{equation}\label{eq:spin-dicke-hamiltonian}
H_{SD}
=
\omega (s + S_{z})
+
\omega_{0} J_{z}
+
\frac{\lambda}{\sqrt{NM}}
(S_{-}+S_{+})
(J_{-}+J_{+}).
\end{equation}
Here, the sizes of the spin operators $(S_{z},S_{\pm})$ and $(J_{z},J_{\pm})$ are $M/2$ and $N/2$, respectively. For brevity, we shall drop this dependence unless mentioned otherwise.

We call the model corresponding to Eq.~\eqref{eq:spin-dicke-hamiltonian} \textit{spin-Dicke} model. 
Here, the bosonic mode in the Dicke model is replaced by a spin-$M/2$ system interacting with the atomic spin-$N/2$ system, whose operators are defined as
$J_{z} = \sum_{n=1}^{N}\sigma_{an}^{z}/2$
and
$J_{\pm} = \sum_{n=1}^{N}\sigma_{an}^{\pm}$.
In Fig.~\ref{fig:avg-boson-num-dicke-and-spin-dicke}(b), we compare the Dicke model with the spin-Dicke model for several values of $M$. 
As the number of qubits representing the bosonic mode increases, the model converges to the Dicke model in terms of the ground-state average 
boson number $\langle a^{\dagger}a\rangle$. 

In Fig.~\ref{fig:avg-boson-num-dicke-and-spin-dicke}(c), we show how the spin-Dicke model approaches its thermodynamic limit as $N$ increases, while the ratio $N/M$ is kept fixed. We derive the average boson number in this limit
as
\begin{equation}\label{eq:avg-adag-a-over-N-sd-thermlim}
    \frac{\langle a^{\dagger}a \rangle}{N}
    \rightarrow 
    \frac{M}{2N} -\frac{M}{2N} \sqrt{1 - \frac{4\frac{N}{M}\frac{\lambda^2}{\omega^2}}{1 + 4\frac{N}{M}\frac{\lambda^2}{\omega^2}}\left[1 - \left(\frac{\lambda_{c}}{\lambda}\right)^{4}\right]},
\end{equation}
for $\lambda > \lambda_{c}=\sqrt{\omega\omega_{0}}/2,$ and ${\langle a^{\dagger}a \rangle}/{N} \rightarrow 0$, otherwise.
The slope jumps from zero to a finite value
\begin{equation}\label{eq:slope-jump-sd}
    \left.\frac{1}{N}\frac{d\langle a^{\dagger}a\rangle}{d(\lambda/\lambda_{c})}\right|_{\lambda=\lambda_{c}+0}
    =
    \frac{\omega_{0}}{\omega + \frac{N}{M}\omega_{0}},
\end{equation}
at the same critical point as the Dicke model,
exhibiting the second-order phase transition for any fixed $N/M$.
As we increase $M$ such that $N/M \rightarrow 0$, the average boson number per particle given in Eq.~(\ref{eq:avg-adag-a-over-N-sd-thermlim}) and the jump in its first derivative at the critical point, Eq.~(\ref{eq:slope-jump-sd}), converge to $[1 - (\lambda_{c}/\lambda)^{4}](\lambda/\omega)^{2}$ and $\omega_{0}/\omega$, respectively,
recovering the thermodynamic limit of the Dicke model.

\section{VQE Implementation}
\begin{figure}
    \centering
\includegraphics[width=\linewidth]{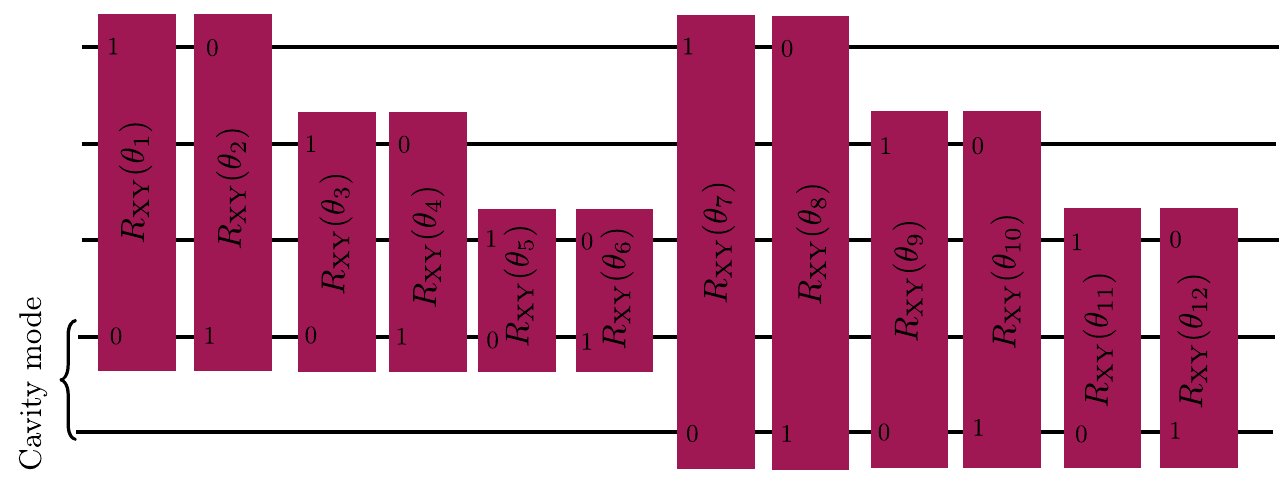}
    \caption{Circuit for the single layer of parameterized symmetry preserving ansatz in the case of $N=3$ and $M=2$, where the first three qubits are qubits that encode the atoms, and the last two qubits encode the cavity mode. Each gate operation takes the form of $\exp[-i\theta X_{a,n}Y_{c,m}]$ or $\exp[-i\theta Y_{a,n}X_{c,m}]$, where $Q_{a,n}$ and $Q_{c,m}$ for $Q \in \{X,Y\}$ represent Pauli operators acting on the $n$th atomic qubit and the $m$th cavity qubit, respectively.}
    \label{fig:ansatz2}
\end{figure}
We now consider the implementation of VQE to evaluate the ground and excited state of the spin-Dicke Hamiltonian  in Eq.~\eqref{eq:spin-dicke-hamiltonian}.
The key step in the VQE implementation pipeline is the selection of an appropriate ansatz, the parameterized quantum circuit whose parameters are optimized variationally to approximate the eigenstates of the target Hamiltonian. The quality of the final solution strongly depends on this choice as the ansatz defines the accessible region of the Hilbert space explored during the optimization. Therefore, the ansatz must be sufficiently expressive to capture the relevant physical states of the system, including both the ground state and low-lying excited states. In the present case, for coupling strengths $\lambda > 0$, the states of interest generally exhibit nontrivial correlations and become increasingly entangled as the coupling increases. As a result, the ansatz must be capable of generating entanglement across the qubits while still remaining shallow enough to be practically implementable. 

 As the model Hamiltonian  possesses parity symmetry, it is often advantageous to design the ansatz so that it respects this symmetry. A symmetry-preserving ansatz restricts the variational search to the physically relevant subspace, which can improve optimization stability and reduce the likelihood of converging to local minima~\cite{gard2020efficient}. Alternatively, one may employ a general ansatz supplemented with a Lagrange-multiplier penalty term in the cost function to enforce the desired symmetry constraints. However, in our case, such symmetry-aware cost functions do not lead to improved performance and are more resource-intensive, as the generic ansatz can increase the variational search space. More detailed discussion on ansatz construction and structure of optimization landscape for various representative cases is given in Sec.~\ref{AnsatzConstruction}.

We design a problem-inspired variational form that directly reflects the interaction structure of the truncated Dicke Hamiltonian. Furthermore, we construct layers based on parametrized two-qubit operations acting pairwise between the cavity and each atomic qubit. Specifically, for each cavity–atom pair, we apply a unitary of the form
\begin{equation}
U(\theta) = \exp \left( - i  \theta Y \otimes X /2 \right).
\end{equation}
This unitary preserves the parity symmetry of the system because its generator, $Y\otimes X$, commutes with the parity operator. Consequently, the evolution generated by $U(\theta)$ remains entirely within a fixed parity sector and does not couple states belonging to different parity subspaces, restricting the variational search to the correct symmetry sector. The ansatz is further designed to respect exchange symmetry between the cavity and atomic subsystems~\cite{gard2020efficient}. The $U(\theta)$ can be implemented using $R_{XY}(\theta)$ gate. Although a single $R_{XY}$ interaction is not invariant under the exchange of the two subsystems, this symmetry is restored by applying two consecutive $R_{XY}$ gates in opposite directions, one acting from the cavity to the atom and the other from the atom to the cavity. Therefore, in practice, we implement this unitary using two consecutive parametrized gates of type $R_{XY}(\theta)$. Thus, two parametrized $R_{XY}(\theta)$ rotations are applied per interaction 
block, ensuring that both interaction directions are symmetrically incorporated. This structure is directly motivated by the physical coupling in the Dicke Hamiltonian and reduces the search space to states that are more physically relevant, thereby improving optimization efficiency compared to fully generic ansätze. The complete problem-inspired ansatz is built by stacking such interaction blocks between $M$ cavity qubits with all $N$ atomic qubits. The corresponding circuit construction for a single problem-inspired ansatz corresponding to $N=3$ and $M=2$ is given in Fig.~\ref{fig:ansatz2}.

We employ the same ansatz for excited-state calculations as well. To target excited states, the ansatz is initialized within an appropriate excitation manifold, allowing the VQE optimization to converge to the lowest-energy eigenstate belonging to that manifold. Unlike the widely used variational quantum deflation (VQD) algorithm~\cite{higgott2019variationalVQD, wen2021variationalVQD}, which determines excited states by adding overlap-penalty terms to exclude previously computed lower-energy states, our approach directly targets symmetry-resolved excitation sectors without requiring explicit deflation. This results in a simpler optimization procedure while naturally exploiting the symmetry structure of the spin-Dicke model.

We emphasize that the symmetry-preserving ansatz introduced here cannot capture spontaneous symmetry breaking or the emergence of symmetry-broken superradiant states in the thermodynamic limit. However, the present work focuses on the finite-size Dicke model, where spontaneous symmetry breaking is absent and the exact eigenstates retain the parity symmetry of the Hamiltonian. Consequently, a symmetry-preserving ansatz is well suited for accurately describing both the ground and excited states of the finite system. Although the results presented in this work are restricted to the resonant case, $(\omega=\omega_{0}=1)$, the symmetry-preserving ansätze developed here are not limited to this parameter regime. They can be applied directly to the spin-Dicke representation of the inhomogeneous Dicke Hamiltonian introduced in Eq.~(\ref{eq:hamiltonian_inhomo}), enabling variational studies of systems with site-dependent atomic transition frequencies and coupling strengths.
\subsection{VQE simulation results}
We simulate the VQE using Qiskit's \texttt{StatevectorSimulator}, which mimicks an ideal quantum computer. We employ the Rotosolve algorithm \cite{Ostaszewski2021structure} for the classical optimization in the VQE calculations. Rotosolve is a gradient-free optimizer that exploits the trigonometric dependence of the VQE cost function on each variational parameter. When all other parameters are held fixed, the cost function mathematically becomes a one-dimensional trigonometric function of the selected parameter, allowing its optimum to be determined analytically. For ansätze composed of Pauli rotation gates, this requires only three cost function evaluations per parameter update. Consequently, a complete optimization sweep over $P$ variational parameters requires approximately $3P$ evaluations of the quantum cost function. Unlike generic derivative-free optimizers such as COBYLA, Nelder–Mead, or Powell, which rely solely on numerical function evaluations, Rotosolve explicitly exploits the analytical structure of the objective function, making it particularly well suited for variational quantum circuits composed of parameterized single- and two-qubit Pauli rotation gates \cite{Ostaszewski2021structure, Wierichs2022generalparameter}. \\
\begin{figure}[h!]
    \centering
\includegraphics[width=\linewidth]{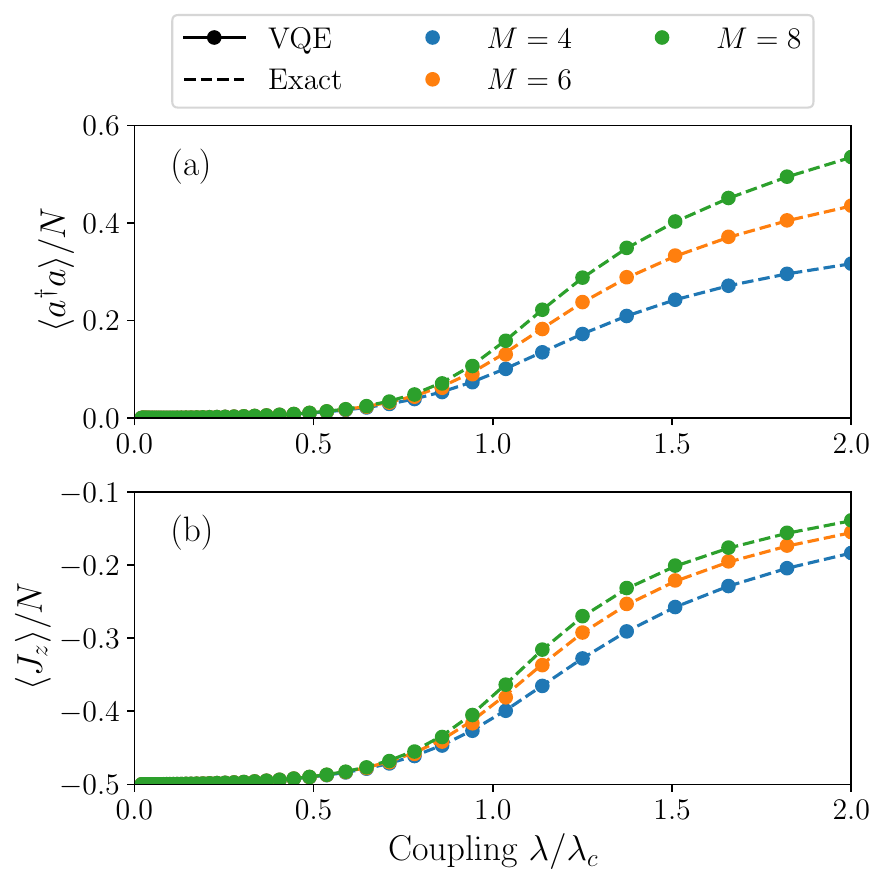}
\caption{Expectation values of the cavity excitations per atom and collective atomic operator $J_z$ per atom for the ground state of the spin-Dicke model. The results obtained from the variational quantum eigensolver (VQE) using the problem-inspired ansatz (dots) are compared with those from exact diagonalization (dashed lines). In this case, the two results coincide very closely. The system parameters are $N=4$ atoms, $\omega =\omega_{0} =  1$ , while an ansatz depth of two layers was used.}
    \label{fig:SZ}
\end{figure}
\begin{figure}
    \centering
    \includegraphics[width=\linewidth]{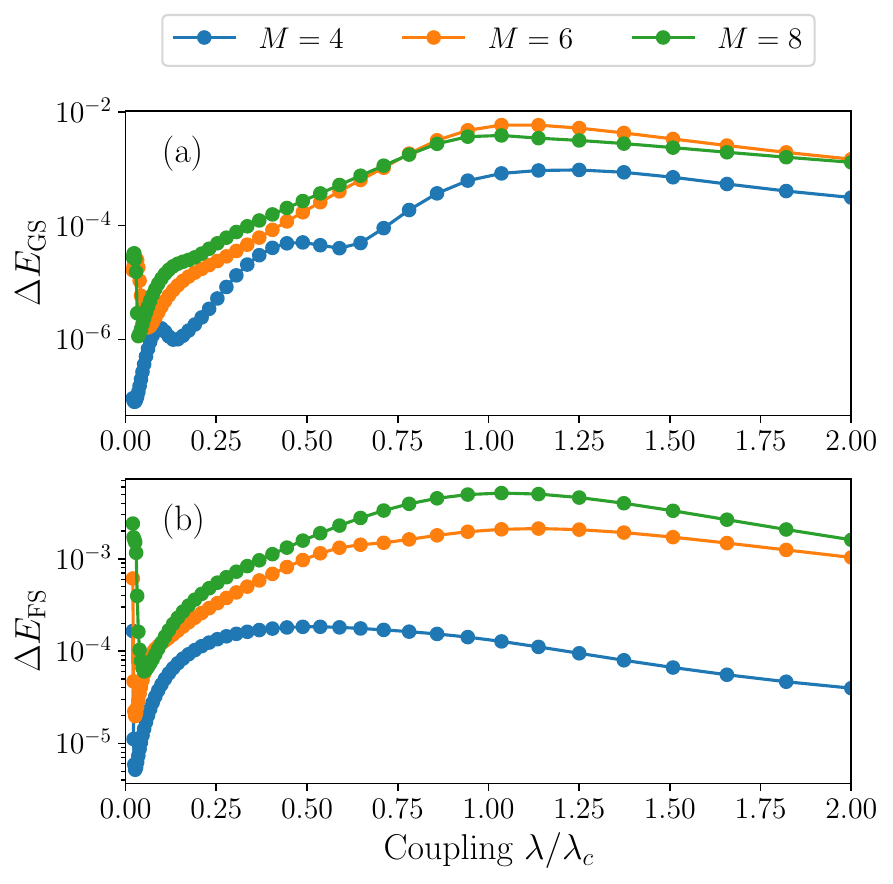}
    \caption{Energy deviation $\Delta E =E_{\text{VQE}} - E_{\text{exact}}$ for the ground state (GS) and first excited state (FS) of the spin-Dicke model, shown in panels~(a) and~(b), respectively. The results are obtained using a problem-inspired ansatz within the variational quantum eigensolver (VQE) and are compared against exact diagonalization. The system parameters are the number of atoms $N=4$, with $\omega = \omega_0 = 1$, and an ansatz depth of two layers.}
    \label{fig:deltaE}
\end{figure}

In Fig.~\ref{fig:SZ}, we compare the expectation values of the cavity excitation number per atom and the collective spin polarization, $\langle J_z\rangle/N$, in the ground state of the spin-Dicke model obtained from VQE with those computed by exact diagonalization. Throughout this section, we consider a system of $N=4$ atoms with $\omega_{0}=\omega=1$ and employ a two-layer symmetry-preserving variational ansatz. The VQE results (dots) closely follow the exact solutions (dashed lines) over the entire range of coupling strengths considered, demonstrating that the proposed VQE ansatz accurately captures the evolution of the relevant order parameters and faithfully reproduces the critical behavior of the finite-size Dicke model. The cavity excitation number per atom is inferred from the expectation value of the collective spin operator, $S_z$, for the cavity encoded qubits allowing for a direct comparison with the corresponding results of the Dicke model. This correspondence between the spin-Dicke model and the Dicke model systematically improves with increasing cavity spin size, $M$, and becomes asymptotically exact in the limit $M \gg N$. Although the proposed VQE framework is directly applicable to larger cavity truncations, the numerical results presented here are restricted to moderate values of $M$. This restriction arises solely from the exponential computational cost of the classical statevector simulations employed for benchmarking, rather than from any fundamental limitation of the variational approach itself.

Fig.~\ref{fig:deltaE} quantifies the accuracy of the variational solutions by showing the energy deviation, $\Delta E=E_{\mathrm{VQE}}-E_{\mathrm{exact}}$, for the ground state [Fig.~\ref{fig:deltaE}(a)] and the first excited state [Fig.~\ref{fig:deltaE}(b)]. For both states, the variational energies remain in excellent agreement with the exact results throughout the parameter regime studied, with typical energy errors of the order of $10^{-2}$. The largest deviations occur in the vicinity of the critical coupling, where the rapid evolution of the eigenstates and the associated increase in quantum correlations make the variational optimization more demanding. Nevertheless, the overall accuracy remains high, and these deviations can be systematically reduced by increasing the expressibility of the ansatz, for example, through additional variational layers.
\section{The spin-Dicke ansatz } \label{AnsatzConstruction}
In this section, we systematically investigate the construction of symmetry-preserving variational ansätze for the spin-Dicke model introduced in Eq.~(\ref{eq:spin-dicke-hamiltonian}). By analyzing its analytical structure and the optimized variational parameters for simple analytically assessable cases, we identify the essential ingredients required to accurately describe the ground state while reducing the number of independent variational parameters. This analysis provides physical insight into the variational manifold, identifies opportunities for reducing the number of variational parameters, and forms the basis for constructing more general ansätze. In Sec.~\ref{sec:ansatz-spin-boson-hybrid-for-dicke}, we extend these ideas to develop a variational ansatz for the ``regular'' Dicke model.

When the coupling is zero, all the spin-$1/2$ particles representing the atoms as well as those representing the bosonic mode are in the down state to minimize the energy. To simulate each particle with a qubit, 
we make correspondence between the spin-down (spin-up) state of each particle and the zero (one) state of each qubit as
\begin{equation*}
\begin{split}
    |{\downarrow}\rangle &\mapsto |0\rangle,\\
    |{\uparrow}\rangle &\mapsto |1\rangle.
\end{split}
\end{equation*}
Thus, the spin-$1/2$ operators of each particle, $\sigma^{z}=|{\uparrow}\rangle\langle {\uparrow}| - |{\downarrow}\rangle\langle {\downarrow}|$ and $\sigma^{x} = |{\downarrow}\rangle\langle {\uparrow}| - |{\uparrow}\rangle\langle {\downarrow}|$, map to the Pauli operators of each qubit,
\begin{equation*}
\begin{split}
    \sigma^{z} 
    &\mapsto |1\rangle\langle 1| - |0\rangle\langle 0| = -Z,\\
    \sigma^{x} &\mapsto |0\rangle\langle 1| + |1\rangle\langle 0| = X,
\end{split}
\end{equation*}
respectively.
From this spin-to-qubit mapping, and dropping the constant $\omega S$, the spin-Dicke Hamiltonian given in Eq.~(\ref{eq:spin-dicke-hamiltonian}) is encoded as
\begin{equation}\label{eq:H-qubits}
    H_{SD} \mapsto H=\omega\left(-\frac{Z_{c}}{2}\right) + \omega_{0}\left(-\frac{Z_{a}}{2}\right) + \frac{\lambda}{\sqrt{NM}}X_{a}X_{c},
\end{equation}
where $Q_{a} = Q_{a1} + \cdots + Q_{aN}$ and $Q_{c} = Q_{c1}+\cdots+Q_{cM}$ are collective Pauli operators for the atomic and cavity qubits, respectively, for $Q \in \{X,Z\}$.
We note that, upon arbitrary exchange among atomic qubits, the collective Pauli operator $Q_{a}$ is invariant.
Same goes for $Q_{c}$ under any permutation between cavity qubits.

We start from the free Hamiltonian, i.e., $\lambda=0$, whose ground state is $|0\rangle_{a}|0\rangle_{c}$, where $|0\rangle_{a}=\otimes_{n=1}^{N}|0\rangle_{an}$ and $|0\rangle_{c}=\otimes_{m=1}^{M}|0\rangle_{cm}$ are the eigenstates of $Z_{a}$ and $Z_{c}$ with eigenvalues $N$ and $M$, respectively. The corresponding energy is 
\begin{equation*}
    E_{G}(\lambda=0)= \omega\left(-\frac{M}{2}\right)+ \omega_{0}\left(-\frac{N}{2}\right).
\end{equation*}
Since the Hamiltonian consists of $Z_{a(c)}$ and $X_{a}X_{c}$, which preserve the number of excitations or flip qubits in pairs,
every non-degenerate eigenstate has definite parity.
In particular, the ground state for $\lambda = 0$ is even since there is no excitation (i.e., all qubits are in their zero states).
Furthermore, the ground state is even for all $\lambda$, as
continuously increasing the coupling strength 
for finite $N$ and $M$ would preserve the parity of the ground state due to the discreteness of the parity.
In the opposite limit, where the coupling strength is large compared to $\omega$ and $\omega_{0}$, the Hamiltonian is dominated by the interaction term, $X_{a}X_{c}$, whose ground state with even parity is
\begin{equation}\label{eq:ground-state-inf-coupling-even-parity}
    \frac{1}{\sqrt{2}}|+\rangle_{a}|-\rangle_{c} + \frac{1}{\sqrt{2}}|-\rangle_{a}|+\rangle_{c},
\end{equation}
where $X_{a}|\pm\rangle_{a} = \pm N|\pm\rangle_{a}$ and $X_{c}|\pm\rangle_{c} = \pm M|\pm\rangle_{c}$.
The corresponding energy asymptotically follows
\begin{equation}\label{eq:energy-inf-coup}
    E_{G}(\lambda)/\lambda \xrightarrow{\lambda \rightarrow \infty} -\sqrt{NM}.
\end{equation}
The variational ansatz should be able to reproduce the ground state in the limit cases, namely $\lambda = 0$ and $\lambda \rightarrow \infty$.

\subsection{Ansatz for $N=1$ and $M=1$}

For $N=1$ and $M=1$, it is possible to prepare the ground state with a single two-qubit gate as
\begin{equation}\label{eq:ansatz-N1-M1}
\begin{split}
    |\theta\rangle 
    &= e^{-i\theta X_{a1}Y_{c1}/2}|0\rangle_{a1}|0\rangle_{c1}\\
    &= \cos\left(\theta/2\right)|0\rangle_{a1}|0\rangle_{c1} + \sin\left(\theta/2\right)|1\rangle_{a1}|1\rangle_{c1},
\end{split}
\end{equation}
where we expand the two-qubit gate using $e^{-i\theta A} = \cos{(\theta)}I -i \sin{(\theta)}A$ for $A^{2}=A=A^{\dagger}$.
This choice is sufficient to generate the ground state, since there are only two states with even parity, namely $|0\rangle_{a1}|0\rangle_{c1}$ and $|1\rangle_{a1}|1\rangle_{c1}$, in the computational basis, and that the coefficients in the two-dimensional, even-parity basis should be real up to a global phase due to the reality of the Hamiltonian matrix in the basis.
To confirm this point, we search for the optimal angle that minimizes the energy expectation value,
\begin{equation*}
    \langle \theta | H |\theta\rangle
    = -\frac{\omega+\omega_{0}}{2}\cos\theta
    + {\lambda}\sin\theta,
\end{equation*}
to get
\begin{equation}\label{eq:opt-angle-N1-M1}
    \tan\theta = -\frac{2\lambda}{\omega + \omega_{0}},
\end{equation}
for which $\langle \theta | H |\theta\rangle$ gives the exact ground-state energy,
\begin{equation*}
    E_{G}(\lambda) = -\sqrt{\lambda^{2} +\left(\frac{\omega+\omega_{0}}{2}\right)^{2}},
\end{equation*}
for $N=1$ and $M=1$.
We note that we have $\theta = 0$ for $\lambda = 0$, so that $|\theta\rangle = |0\rangle_{a1}|0\rangle_{c1}$,
whereas in the infinite-coupling limit, the angle and the ansatz converge to $\pi/2$ and $[|0\rangle_{a1}|0\rangle_{c1} + |1\rangle_{a1}|1\rangle_{c1}]/\sqrt{2}$, which is identified as the asymptotic ground state given in Eq.~(\ref{eq:ground-state-inf-coupling-even-parity}) via $|\pm\rangle_{a(c)} = |\pm\rangle_{a1(c1)} = [|0\rangle_{a1(c1)} \pm |1\rangle_{a1(c1)}]/\sqrt{2}$ for $N=M=1$.
We also note that we can also use $\exp[-i\theta Y_{a1}X_{c1}/2]$ instead of Eq.~(\ref{eq:ansatz-N1-M1}) to obtain the same ansatz. 

Our choice of using $X_{a1}Y_{c1}$ gate shall be further justified as we investigate the ansatz for multiple qubits in the following section.

\subsection{Ansatz for $N=1$ and $M\ge1$}\label{sec:ansatz-N1-M}

We now extend the ansatz to have multiple cavity qubits ($M>1$). Since the model is invariant under any exchange between the cavity qubits, we generalize Eq.~(\ref{eq:ansatz-N1-M1}) by adding the two-qubit gate for each cavity qubit with the same angle as
\begin{equation}\label{eq:ansatz-N1-M}
    |\theta\rangle 
    = \underbrace{e^{-i\theta X_{a1}Y_{cM}/2}\cdots e^{-i\theta X_{a1}Y_{c1}/2}}_{e^{-i\theta X_{a1}Y_{c}/2}}
    |0\rangle_{a1}|0\rangle_{c},
\end{equation}
where $Y_{c} = Y_{c1}+\cdots+Y_{cM}$. 
Below,
we show a quantum circuit for preparing this ansatz, where each two-qubit gate represents $\exp[-i\theta X_{a1}Y_{cm}/2]$ for $1 \le m \le M$:

\begin{equation*}
    \Qcircuit @C=1em @R=.8em @!R { 
        & \lstick{c1} & \sgate{Y}{4} & \qw & \qw & \qw & \qw
        & & &
        & & &
        \\
        & \lstick{c2} & \qw & \sgate{Y}{3} & \qw & \qw & \qw 
        & & &
        & \dstick{M} & &
        \\
        & \lstick{\vdots} & & & \ddots & & 
        & \push{\equiv} & &
        \lstick{c} & {/} \qw & \sgate{Y}{2} & \qw
        \\
        & \lstick{cM} & \qw & \qw & \qw & \sgate{Y}{1} & \qw 
        & & &
        & & &
        \\
        & \lstick{a1} & \gate{X} & \gate{X} & \ustick{\cdots} \qw & \gate{X} & \qw 
        & & &
        \lstick{a1} & \qw & \gate{X} & \qw
        \\
        & & \ustick{\theta} & \ustick{\theta} & & \ustick{\theta} &
        & & &
        & & \ustick{\theta} &
    }
\end{equation*}
The ansatz can reach the asymptotic ground state given in Eq.~(\ref{eq:ground-state-inf-coupling-even-parity}) for the infinite-coupling limit. This is from the fact that $|0\rangle_{a1} = [|+\rangle_{a1}+|-\rangle_{a1}]/\sqrt{2}$ is the equal superposition of the eigenstates of $X_{a1}$ so that 
\begin{equation*}
    |\theta\rangle 
    = \frac{1}{\sqrt{2}}\sum_{\mu=\pm}|\mu\rangle_{a1} \otimes e^{-i\mu\theta Y_{c}/2}|0\rangle_{c},
\end{equation*}
and that $e^{-i \theta Y_{c}/2}$ rotates the state $|0\rangle_{c}$ to $|\pm\rangle_{c}$ when $\theta = \pm\pi/2$. Thus, choosing the correct sign, the ansatz can reach the asymptotic limit as $\theta \rightarrow -\pi/2$.
We note that if we select $\exp[-i\theta Y_{a1}X_{cm}/2]$ as the building block for constructing the ansatz, instead of Eq.~(\ref{eq:ansatz-N1-M}), this limit is not met. However, this ansatz can be used under the opposite conditions where $N > 1$ and $M=1$.


The energy expectation value of the ansatz, Eq.~(\ref{eq:ansatz-N1-M}), can be computed as (Appendix~\ref{sec:single-layer-energy-expectation-value})
\begin{equation}\label{eq:avg-H-ansatz-N1-M}
    \langle \theta | H |\theta\rangle
    = -\frac{\omega}{2} M\cos\theta
    -\frac{\omega_{0}}{2} \cos^{M}\theta
    + {\lambda} \sqrt{M}\sin\theta.
\end{equation}
For $\lambda = 0$, the expectation value can reach the ground-state energy if $\cos\theta=-1$. When $\lambda/\lambda_{c} \rightarrow \infty$, the interaction term becomes dominant and the minimal energy can be reached if $\sin\theta = -1$. 

\begin{figure}
    \centering
    \includegraphics[width=\linewidth]{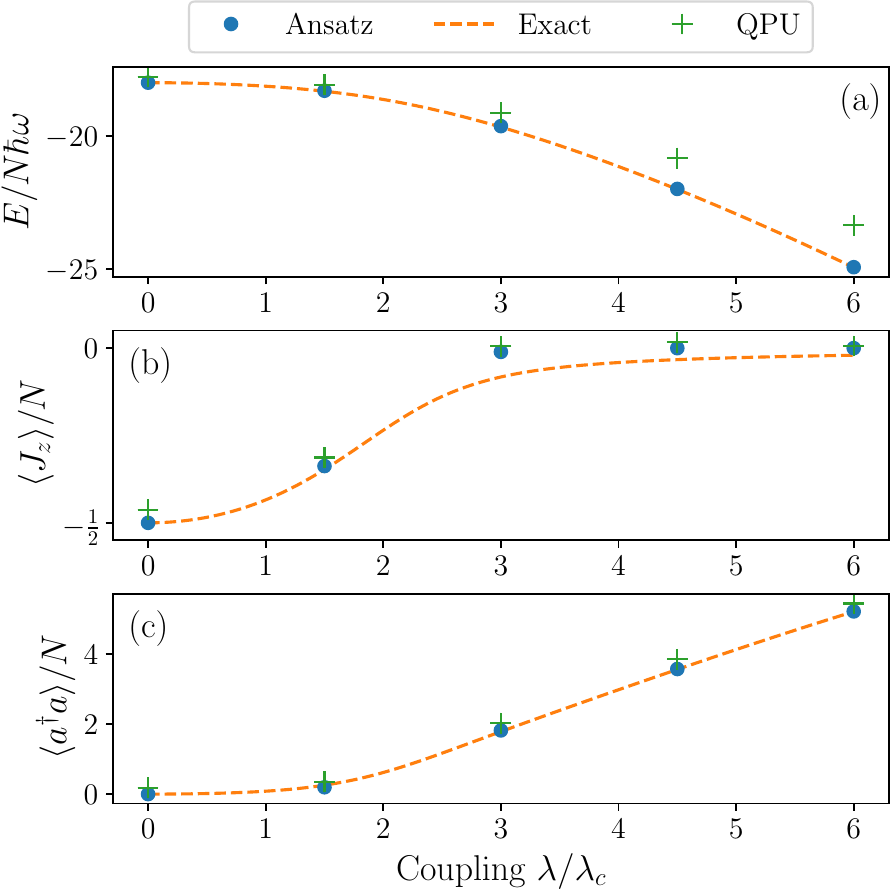}
    \caption{Optimization results for the spin-Dicke model with $N=1$ and $M=35$. The orange dashed linies denote the exact ground-state expectation values obtained from exact diagonalization of the spin-Dicke Hamiltonian. The blue circles show the expectation values evaluated using the variational ansatz in Eq.~(\ref{eq:ansatz-N1-M}) with classically optimized parameters. The green crosses correspond to measurements performed on a trapped-ion quantum processing unit (QPU), where the ansatz state was prepared using the same classically optimized angle (see Appendix~\ref{sec:qpu}).
    }
    \label{fig:spin-dicke-N1-multiM}
\end{figure}

For each value of the coupling strength, we first determine the optimal angle by classically minimizing the energy expectation value of the ansatz. The optimized parameter is then used to prepare the variational state, and the corresponding expectation values are compared with the exact results obtained from exact diagonalization of the spin-Dicke Hamiltonian.
Fig.~\ref{fig:spin-dicke-N1-multiM} presents these results for the spin-Dicke model with a single atomic qubit and $M=35$ cavity qubits. To assess the experimental feasibility of the proposed ansatz, we also prepare the classically optimized variational state on a trapped-ion quantum processing unit (QPU) and measure the corresponding expectation values (see Appendix~\ref{sec:qpu} for the details of the QPU simulation). We emphasize that the variational optimization itself is performed entirely classically using the energy expectation value given in Eq.~(\ref{eq:avg-H-ansatz-N1-M}); the QPU is used only for state preparation and measurement, rather than for executing the complete VQE optimization loop.
 
%


\subsection{Ansatz for $N\ge1$ and $M\ge1$}\label{sec:ansatz-N-M}
So far, we have considered a single atomic qubit coupled to multiple cavity qubits. 
If we add one or more atomic qubits, we would connect the new qubits with the cavity qubits with the same gate set to preserve the exchange symmetry of the model.
Thus, for $N>1$, we may extend the single-atom ansatz given in Eq.~(\ref{eq:ansatz-N1-M}) as
\begin{equation*}
    |\theta\rangle 
    = \underbrace{e^{-i\theta X_{aN}Y_{c}/2} \cdots e^{-i\theta X_{a1}Y_{c}/2}}_{e^{-i\theta X_{a}Y_{c}/2}}|0\rangle_{a}|0\rangle_{c},
\end{equation*}
where the quantum circuit of the collective gates can be depicted by the following diagram:
\begin{equation*}
    \Qcircuit @C=1em @R=.8em @!R { 
        & & \dstick{M} & \dstick{\theta} & \dstick{\theta} & & \dstick{\theta} & 
        & & & 
        & \dstick{M} & \dstick{\theta} &
        \\
        & \lstick{c} 
        & {/} \qw & \sgate{Y}{1} & \sgate{Y}{2} & \dstick{\cdots} \qw & \sgate{Y}{4} & \qw 
        & & & 
        \lstick{c} & {/} \qw & \sgate{Y}{3} & \qw
        \\
        & \lstick{a1} 
        & \qw & \gate{X} & \qw & \qw & \qw & \qw 
        & & & 
        & & &
        \\
        & \lstick{a2} 
        & \qw & \qw & \gate{X} & \qw & \qw & \qw 
        & & & 
        & \dstick{N} & &
        \\
        & \lstick{\vdots} 
        & & & & \ddots & & 
        & \push{\equiv} & & 
        \lstick{a} & {/} \qw & \gate{X} & \qw
        \\
        & \lstick{aN} 
        & \qw & \qw & \qw & \qw & \gate{X} & \qw 
        & & & 
        & & &
        \\
        \\
    }
\end{equation*}
However, this ansatz does not reach the infinite-coupling ground state, Eq.~(\ref{eq:ground-state-inf-coupling-even-parity}), for which the interaction term of the Hamiltonian is given by $\langle X_{a}X_{c}\rangle = -NM$, resulting in Eq.~(\ref{eq:energy-inf-coup}).
To see this, we show $\langle \theta | X_{a}X_{c}|\theta \rangle > -NM$.
From the expression of the expectation value with respect to the ansatz given in Eq.~(\ref{eq:avg-theta-ket}) for $N>1$,
\begin{equation*}
    \langle \theta | X_{a}X_{c} | \theta \rangle 
    = NM\cos^{N-1}\theta\sin\theta
    > -NM,
\end{equation*}
we see that for any $\theta$, the expectation value is strictly above that of the infinite-coupling ground state.

Instead of starting from $|0\rangle_{a}|0\rangle_{c}$ at $\theta = 0$,
we consider an unknown state $|\phi\rangle_{a}$ combined with $|0\rangle_{c}$ to construct a generalized ansatz of the form 
\begin{equation}\label{eq:theta-phi-ket}
    e^{-i\theta X_{a}Y_{c}/2}|\phi\rangle_{a}|0\rangle_{c}.
\end{equation}
When $N=1$, the ansatz reduces to Eq.~(\ref{eq:ansatz-N1-M}) for general $M$ by defining $|\phi\rangle_{a} = |0\rangle_{a} = |0\rangle_{a1}$.
To derive conditions on $\theta$ and $|\phi\rangle_{a}$ for $N>1$ to reach the infinite-coupling ground state given in Eq.~(\ref{eq:ground-state-inf-coupling-even-parity}), we equate the ansatz with the ground state as
\begin{equation*}
    e^{-i\theta X_{a}Y_{c}/2}|\phi\rangle_{a}|0\rangle_{c}
    \stackrel{!}{=} 
    \frac{1}{\sqrt{2}}|+\rangle_{a}|-\rangle_{c} + \frac{1}{\sqrt{2}}|-\rangle_{a}|+\rangle_{c}
\end{equation*}
for general $M$ and $N$.

\begin{figure}
    \centering
    \includegraphics[width=\linewidth]{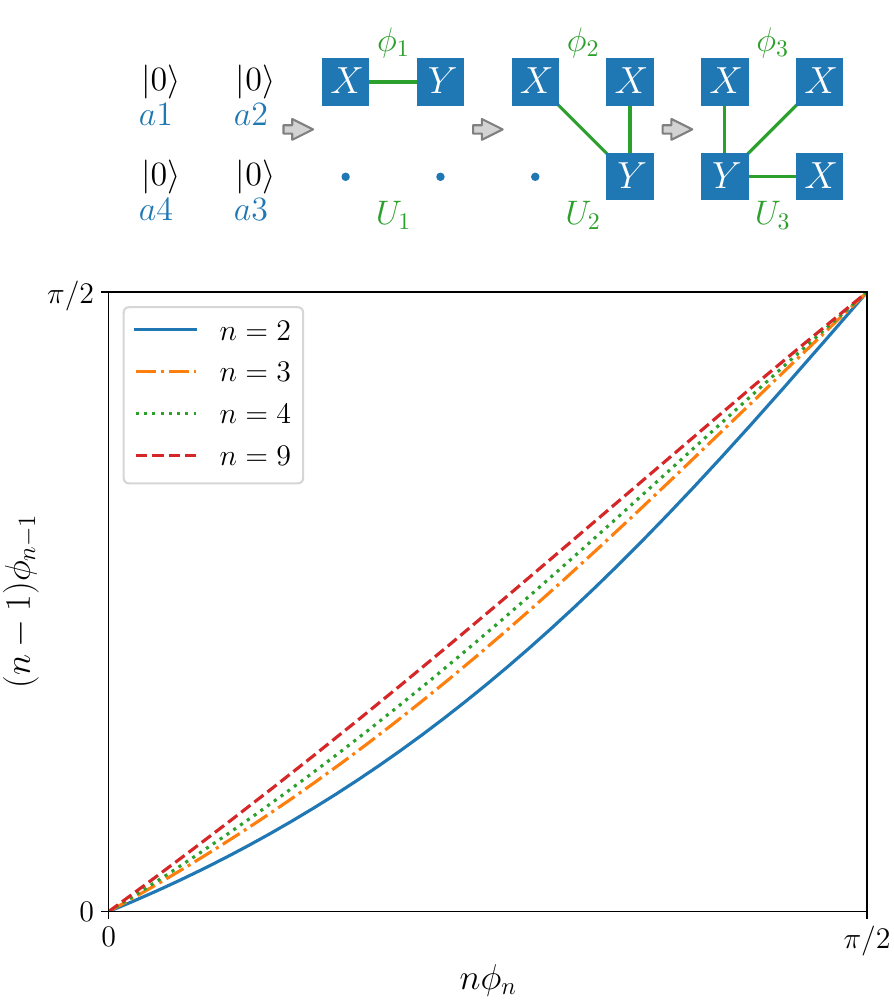}
    \caption{On the upper panel, we show part of the ansatz preparing the atomic state,Eq.~(\ref{eq:phi1-to-phi-Nminus1-ket-a}),for $N=4$. For a given $\phi_{3} = \phi$, the generated four-qubit state can be exchange symmetric by setting $\phi_{2}$ and $\phi_{1}$ according to Eqs.~(\ref{eq:tan-phi1-half-for-given-phi2}) and (\ref{eq:tan-phi2-for-given-phi3}), respectively. On the lower panel, we show $\phi_{2}$ and $\phi_{1}$ as functions of $\phi_{3}$ and $\phi_{2}$, respectively, as well as those with some higher values of $n$.
    }
    \label{fig:atomic-ansatz}
\end{figure}

Noting that $X_{a}|\pm\rangle_{a} = \pm N|\pm\rangle_{a}$, we invert $e^{-i\theta X_{a}Y_{c}/2}$ on the left-hand side to get
\begin{equation*}
    |\phi\rangle_{a}|0\rangle_{c}
    \stackrel{!}{=} 
    \frac{1}{\sqrt{2}}e^{i\theta NY_{c}/2}|+\rangle_{a}|-\rangle_{c} 
    + \frac{1}{\sqrt{2}}e^{-i\theta NY_{c}/2}|-\rangle_{a}|+\rangle_{c}.
\end{equation*}
For the right-hand side to be the product state on the left-hand side, we need
\begin{equation}\label{eq:exp-mp-i-theta-N-Yc-half-pm-ket-c-equals-0-ket-c}
    e^{\mp i\theta NY_{c}/2}|\pm\rangle_{c} \stackrel{!}{=} |0\rangle_{c}.
\end{equation}
Identifying $R_{Y_{c}}(\theta) \equiv e^{- i\theta Y_{c}/2}$ as a rotational operator and noting that $R_{Y_{c}}(\pi/2)|0\rangle_{c} = |+\rangle_{c}$,
the smallest angle in magnitude satisfying Eq.~(\ref{eq:exp-mp-i-theta-N-Yc-half-pm-ket-c-equals-0-ket-c}) is 
\begin{equation}\label{eq:theta-asympt-inf-coup}
    \theta = -\pi/2N. 
\end{equation}
In general, the angle subject to
    $\sin{(N\theta)} = -1$
gives Eq.~(\ref{eq:exp-mp-i-theta-N-Yc-half-pm-ket-c-equals-0-ket-c}) up to a global phase.
For the atomic part, we are left with a Greenberger–Horne–Zeilinger state \cite{greenberger1989going} in the $X$ basis,
\begin{equation}\label{eq:phi-ket-a-inf-coup}
    |\phi\rangle_{a} 
    \stackrel{!}{=} 
    \frac{1}{\sqrt{2}}|+\rangle_{a}
    + \frac{1}{\sqrt{2}}|-\rangle_{a}.
\end{equation}
For $N=1$, this state is merely the qubit ground state, $|0\rangle_{a1}=(1/\sqrt{2})\sum_{\mu=\pm}|\mu\rangle_{a1}$, which we have used to get the single-atom ansatz, Eq.~(\ref{eq:ansatz-N1-M}), encompassing the infinite-coupling ground state. For $N=2$, we can generate the maximally entangled state via a single MS gate as
\begin{equation*}
    e^{-i\phi Y_{a2}X_{a1}/2}|0\rangle_{a2}|0\rangle_{a1}
    = \frac{1}{\sqrt{2}}\sum_{\mu=\pm}e^{-i\phi \mu Y_{a2}/2}|0\rangle_{a2}|\mu\rangle_{a1},
\end{equation*}
which becomes the target state, Eq.~(\ref{eq:phi-ket-a-inf-coup}), when $\phi = \pi/2$, since $|0\rangle_{a2}$ is rotated by $\pm\pi/2$ depending on the sign of $|\pm\rangle_{a2}$. For general $N$, the maximally entangled state can be generated via $N-1$ layers of gates,
\begin{equation}\label{eq:phi1-to-phi-Nminus1-ket-a}
    |{\phi_{1},...,\phi_{N-1}}\rangle_{a} = U_{N-1}(\phi_{N-1})\cdots U_{1}(\phi_{1})|0\rangle_{a},
\end{equation}
where the $n$th layer consisting of $n$ MS gates can be written in terms of $X_{\Sigma{n}} \equiv X_{1}+\cdots X_{n}$ and $Y_{n+1} \equiv Y_{a,n+1}$ as
\begin{equation*}
    U_{n}(\phi_{n}) = e^{-i\phi_{n}Y_{n+1}X_{\Sigma{n}}/2}
    = e^{-i\phi_{n}Y_{n+1}X_{an}/2} \cdots e^{-i\phi_{n}Y_{n+1}X_{a1}/2}.
\end{equation*}
As an example, each layer is visualized for $N=4$ in Fig.~\ref{fig:atomic-ansatz}.
Here, we inductively derive the correct parameters $\{\phi_{n}\}$ of the layers. Suppose we have generated the target state of $n$ qubits, which we denote here as $(1/\sqrt{2})\sum_{\mu=\pm}|\mu\rangle_{\Sigma{n}}$, an equal superposition of the two eigenstates, $|\pm\rangle_{\Sigma{n}} = |\pm\rangle_{n}\cdots|\pm\rangle_{1}$, of $X_{\Sigma{n}}$ with eigenvalues $\pm n$. 
Then, we prepare the $(n+1)$th qubit in its $|0\rangle_{n+1}$ state and apply $U_{n}(\phi_{n})$ to get
\begin{equation*}
    \frac{1}{\sqrt{2}}\sum_{\mu=\pm}
    e^{-i(\mu n \phi_{n})Y_{n+1}/2}|0\rangle_{n+1}|\mu\rangle_{\Sigma{n}}
    =
    \frac{1}{\sqrt{2}}\sum_{\mu=\pm}|\mu\rangle_{n+1}|\mu\rangle_{\Sigma{n}},
\end{equation*}
where we set
\begin{equation}\label{eq:phi-n-inf-coup}
    n \phi_{n} = \pi/2.
\end{equation}
We note that to generate the GHZ state,
the total number of gates are distributed to $N-1$ layers as
\begin{equation*}
    1 + \cdots + (N-1) = \frac{N(N-1)}{2} = \begin{pmatrix}
        N \\ 2
    \end{pmatrix}.
\end{equation*}

So far, we show that the ansatz $|{\theta;\phi}\rangle$ given in Eq.~(\ref{eq:theta-phi-ket}) can express the infinite-coupling ground state, by preparing the initial atomic state $|\phi\rangle_{a}$ to be the GHZ state in the $X_{a}$ basis via Eq.~(\ref{eq:phi1-to-phi-Nminus1-ket-a}). For the preparation of $|\phi\rangle_{a}$, we use $N-1$ angles, $\{\phi_{n}: 1 \le n \le N-1\}$, which we know for the infinite-coupling limit as well as for zero coupling.
For a finite coupling strength, $0 < \lambda < \infty$, we may find optimal angles by minimizing the energy expectation value with respect to the $N-1$ angles and $\theta$. However, as the number of angles scales with the increasing system size, the optimization process incurs numerical overhead which can consume considerable resources, potentially prohibiting scalability of the approach. In the following paragraph, we utilize the exchange symmetry among atomic qubits of the model Hamiltonian given in Eq.~(\ref{eq:H-qubits}), to derive relations between angles of adjacent layers, $\phi_{n}$ and $\phi_{n+1}$.

For $N=2$, there is only $N-1=1$ angle, namely $\phi_{1}$, and the exchange symmetry is kept for any value of the angle since
\begin{equation*}
    U_{1}(\phi_{1})|0\rangle_{a2}|0\rangle_{a1} = \cos\frac{\phi_{1}}{2}|0\rangle_{a2}|0\rangle_{a1}
    + \sin\frac{\phi_{1}}{2}|1\rangle_{a2}|1\rangle_{a1}.
\end{equation*}
We note that $|0\rangle_{a2}|0\rangle_{a1}$ and $|1\rangle_{a2}|1\rangle_{a1}$ are the eigenstates of $Z_{\Sigma{2}} \equiv Z_{a1} + Z_{a2}$ with eigenvalues $+2$ and $-2$, respectively.
To extend the ansatz for $N>2$, 
let us denote the exchange-symmetric eigenstate of $Z_{\Sigma{n}} = Z_{an}+\cdots+Z_{a1}$ with eigenvalue $n - 2m$ as $|m\rangle_{\Sigma{n}}$, which is a Dicke state. 
For example, we have $|0\rangle_{\Sigma{2}} = |0\rangle_{a2}|0\rangle_{a1}$ and $|1\rangle_{\Sigma{2}} = (1/\sqrt{2})(|0\rangle_{a2}|1\rangle_{a1} + |1\rangle_{a2}|0\rangle_{a1})$.
In general, the Dicke state can be written explicitly as
\begin{equation*}
    |m\rangle_{\Sigma{n}} = \begin{pmatrix}
        n \\ m
    \end{pmatrix}^{-\frac{1}{2}}
    \underset{m_{1}+\cdots+m_{n}=m}{\sum_{m_{1}=0}^{1}\cdots\sum_{m_{n}=0}^{1}}
    |{m_{n}}\rangle_{an}\cdots|m_{1}\rangle_{a1}.
\end{equation*}
Writing $|\phi_{1}\rangle_{a}$ in the Dicke basis with 
coefficients $a_{nm} \equiv {}_{\Sigma{n}}\langle m|{\phi_{1},...,\phi_{n-1}}\rangle_{a}$ 
and adding the second layer, we have
\begin{equation*}
    |{\phi_{1},\phi_{2}}\rangle_{a}
    = U_{2}(\phi_{2})|0\rangle_{3}\left[a_{20}|0\rangle_{\Sigma{2}} + a_{22}|2\rangle_{\Sigma{2}}\right],
\end{equation*}
where $a_{20} = \cos(\phi_{1}/2)$ and $a_{22} = \sin(\phi_{1}/2)$.
In general, the state is not exchange-symmetric. For example, when $\phi_{2}=0$ and $0 < \phi_{1} < 2\pi$, swapping the first and the third qubit gives a different state. In Appendix~\ref{sec:N-M-ansatz}, we show that setting $\phi_{1}$ as
\begin{equation}\label{eq:tan-phi1-half-for-given-phi2}
    \tan{\frac{\phi_{1}}{2}} = \frac{1 + \tan{(\phi_{2}/2)}}{1 - \tan{(\phi_{2}/2)}}\tan{\frac{\phi_{2}}{2}}
\end{equation}
for a given $\phi_{2}$
lets the atomic state invariant under an arbitrary permutation among qubits, cf. Eq.~(\ref{eq:tan-phi1-half-of-phi2}).
Since each angle is expected to be bounded by
\begin{equation}\label{eq:phi-n-range}
    0 \le \phi_{n} < \frac{\pi}{2n}
\end{equation}
for $0 \le \lambda < \infty$, following from Eq.~(\ref{eq:phi-n-inf-coup}), the denominator is nonzero.
For $\phi_{2}$ satisfying Eq.~(\ref{eq:phi-n-range}) and $\phi_{1}$ given by Eq.~(\ref{eq:tan-phi1-half-for-given-phi2}), the two-angle ansatz can be written in the exchange-symmetric form as
\begin{equation}\label{eq:phi1-of-phi2-and-phi2-ket-a}
    |{\phi_{1}(\phi_{2}),\phi_{2}}\rangle_{a}
    \equiv a_{30}(\phi_{2})|0\rangle_{\Sigma{3}} + a_{32}(\phi_{2})|2\rangle_{\Sigma{3}}.
\end{equation}
Although there are two angles, $\phi_{1}$ and $\phi_{2}$,
we only need to vary $\phi_{2}$, apart from $\theta$, when optimizing the energy expectation value since $\phi_{1} = \phi_{1}(\phi_{2})$ is determined once $\phi_{2}$ is given.
Similarly, 
for $N = 4$, 
setting $\phi_{2}$ such that
\begin{equation}\label{eq:tan-phi2-for-given-phi3}
    \tan{\phi_{2}} = \frac{\sin{\phi_{3}}}{1 - \sin{\phi_{3}}}
\end{equation}
and $\phi_{1}$ as Eq.~(\ref{eq:tan-phi1-half-for-given-phi2})
ensure the exchange symmetry of the first four atomic qubits
as
\begin{equation}\label{eq:phi3-Sigma4-symm}
\begin{split}
    |\phi_{3}\rangle_{\Sigma{4}}^{\mathrm{symm}}
    &\equiv 
    |{\phi_{1}(\phi_{2}(\phi_{3})),\phi_{2}(\phi_{3}),\phi_{3}}\rangle_{a}\\
    &\equiv \sum_{p=0}^{2}a_{4,2p}(\phi_{3})|2p\rangle_{\Sigma{4}}.
\end{split}
\end{equation}
We note that the coefficients are functions of $\phi_{1}$, $\phi_{2}$ and $\phi_{3}$, but since the first is determined by the second angle, and the second by the third angle, all coefficients are in effect functions solely of $\phi_{3}$, which we denote here as $a_{4,2p}(\phi_{3}) \equiv {}_{\Sigma{4}}\langle{2p}|\phi_{3}\rangle_{\Sigma{4}}^{\mathrm{symm}}$.
Without the constraint on each pair of angles, the ansatz state can have components other than the symmetric basis states, $|0\rangle_{\Sigma{4}}$, $|2\rangle_{\Sigma{4}}$ and $|4\rangle_{\Sigma{4}}$.

For $N \le 4$, we have shown that the ansatz state can be parameterized by a single angle, e.g., $\phi_{3}$ for $N=4$.
To fulfill the exchange symmetry among $N > 4$ atomic qubits, we need more than one degree of freedom. 
For instance, there are two conditions to be satisfied in order to get a symmetric state of $N=5$ atomic qubits, requiring two degrees of freedom.
Below, we show explicitly the two conditions and present a way to 
determine the optimal angle of the ansatz.
Applying the fourth layer on a general four-qubit symmetric state
as
\begin{equation}\label{eq:phi4-Sigma5}
    |\phi_{4}\rangle_{\Sigma{5}}
    \equiv 
    U_{4}(\phi_{4})|0\rangle_{5}[a_{0}|0\rangle_{\Sigma{4}} + a_{2}|2\rangle_{\Sigma{4}}
    + a_{4}|4\rangle_{\Sigma{4}}]
\end{equation}
for three arbitrary real numbers with the normalization condition, $\sum_{p=0}^{2}a_{2p}^{2} = 1$, we use 
\begin{equation*}
    U_{n}(\phi_{n}) = \cos{(\phi_{n}X_{\Sigma{n}}/2)} -iY_{n+1}\sin{(\phi_{n}X_{\Sigma{n}}/2)}
\end{equation*} 
and $-iY_{n+1}|0\rangle_{n+1} = |1\rangle_{n+1}$
to get
\begin{equation}\label{eq:phi-4-Sigma5-expanded}
\begin{split}
    |\phi_{4}\rangle_{\Sigma{5}}
    = \alpha_{0}|0\rangle_{5}|0\rangle_{\Sigma{4}}
    &+ \beta_{1}|1\rangle_{5}|1\rangle_{\Sigma{4}} 
    + \alpha_{2}|0\rangle_{5}|2\rangle_{\Sigma{4}}\\
    &+ \beta_{3}|1\rangle_{5}|3\rangle_{\Sigma{4}} 
    + \alpha_{4}|0\rangle_{5}|4\rangle_{\Sigma{4}},
\end{split}
\end{equation}
where the coefficients
\begin{equation}\label{eq:alpha-m-phi4-beta-m-phi4}
\begin{split}
    \alpha_{m}(\phi_{4};\{a_{2p}\}) 
    &\equiv \sum_{p=0}^{2}{}_{\Sigma{4}}\langle m|\cos\left(\phi_{4}\frac{X_{\Sigma{4}}}{2}\right)|2p\rangle_{\Sigma{4}}a_{2p},\\
    \beta_{m}(\phi_{4};\{a_{2p}\})
    &\equiv \sum_{p=0}^{2}{}_{\Sigma{4}}\langle {m}|\sin\left(\phi_{4}\frac{X_{\Sigma{4}}}{2}\right)|2p\rangle_{\Sigma{4}}a_{2p}
\end{split}
\end{equation}
depend on the initial amplitudes, $a_{0}$, $a_{2}$ and $a_{4}$, given in Eq.~(\ref{eq:phi4-Sigma5}) as well as the angle $\phi_{4}$.
The first term in Eq.~(\ref{eq:phi-4-Sigma5-expanded}) is already symmetric since $|0\rangle_{5}|0\rangle_{\Sigma{4}} = |0\rangle_{\Sigma{5}}$.
For the whole state $|\phi_{4}\rangle_{\Sigma{5}}$ to be symmetric, 
we set the ratios between the coefficients as $\beta_{1}/\alpha_{2} = \sqrt{2/3}$ and $\beta_{3}/\alpha_{4} = \sqrt{4/1}$, to construct
\begin{equation*}
\begin{split}
    |2q\rangle_{\Sigma{5}} 
    &= \sqrt{\frac{2q}{5}}
    |1\rangle_{5}|{2q-1}\rangle_{\Sigma{4}}
    + \sqrt{\frac{5-2q}{5}}
    |0\rangle_{5}|2q\rangle_{\Sigma{4}}
\end{split}
\end{equation*}
for $1 \le q \le 2$.
We note that the two conditions on the ratios can be written as 
\begin{equation*}
    \sqrt{5-2q}\beta_{2q-1} - \sqrt{2q}\alpha_{2q} = 0
\end{equation*}
for $q \in \{1,2\}$, which is a system of linear equations of the initial amplitudes $\{a_{2p}:0 \le p \le 2\}$
since the coefficients are given as Eq.~(\ref{eq:alpha-m-phi4-beta-m-phi4}).
The three normalized amplitudes are determined up to a global sign by solving the linear system for a given $\phi_{4}$.
Let us denote the solution as $a_{2p}^{0}(\phi_{4})$ with the sign fixed by $a_{0}^{0}(0) = 1$, and write the resulting state as
\begin{equation*}
    |\phi_{4}\rangle_{\Sigma{4}}^{0} \equiv 
    \sum_{p=0}^{2}a_{2p}^{0}(\phi_{4})|2p\rangle_{2p}.
\end{equation*}
Note that we do not need information on the previous angles, $\phi_{n}$ for $1 \le n \le 3$, to determine the coefficients.
Thus, if we can prepare the initial four-qubit state in Eq.~(\ref{eq:phi4-Sigma5}) with $a_{2p} = a_{2p}^{0}(\phi_{4})$, the conditions on ratios $\beta_{2q-1}/\alpha_{2q} = \sqrt{2q/(5-2q)}$ are fulfilled so that 
$|\phi_{4}\rangle_{\Sigma{5}}$ given in Eq.~(\ref{eq:phi-4-Sigma5-expanded}) becomes exchange symmetric as
\begin{equation*}
\begin{split}
    |\phi_{4}\rangle_{\Sigma{5}}^{\mathrm{symm}}
    &\equiv U_{4}(\phi_{4})|0\rangle_{5}|\phi_{4}\rangle_{\Sigma{4}}^{0}\\
    &\equiv a_{50}(\phi_{4})|0\rangle_{\Sigma{5}}
    + a_{52}(\phi_{4})|2\rangle_{\Sigma{5}}
    + a_{54}(\phi_{4})|4\rangle_{\Sigma{5}}.
\end{split}
\end{equation*}
However, the circuit preparing the four-qubit state has only one degree of freedom in its coefficients as in Eq.~(\ref{eq:phi3-Sigma4-symm}), namely $a_{40}(\phi_{3})$, $a_{42}(\phi_{3})$, and $a_{44}(\phi_{3})$, there can be some discrepancy from the necessary coefficients, $\{a_{2p}^{0}(\phi_{4})\}$, quantified by the corresponding infidelity, $1 - F(\phi_{4},\phi_{3})$, where
\begin{equation*}
    F_{4}(\phi_{4},\phi_{3}) 
    \equiv \prescript{0}{\Sigma{4}}{\langle \phi_{4}|}\phi_{3}\rangle_{\Sigma{4}}^{\mathrm{symm}}
    = \sum_{p=0}^{2}a_{2p}^{0}(\phi_{4})a_{4,2p}(\phi_{3}).
\end{equation*}
The third angle $\phi_{3}$ is determined by minimizing the infidelity for a given $\phi_{4}$.

For general $n \ge 1$, we compute a set of normalized coefficients, $\{a_{2p}(\phi_{n}):0 \le p \le \lfloor{n/2}\rfloor\}$, for a given $\phi_{n}$ such that
\begin{equation*}
    |\phi_{n}\rangle_{\Sigma,n+1} 
    \equiv U_{n}(\phi_{n})|0\rangle_{n+1} 
    \sum_{p=0}^{\lfloor{n/2}\rfloor}a_{2p}(\phi_{n})|2p\rangle_{\Sigma{n}}
\end{equation*}
is symmetric. Let us denote such coefficients as $a_{2p}^{0}(\phi_{n})$, with the sign degree of freedom fixed via $a_{0}^{0}(0) = 1$ as before.
The corresponding symmetric state reads
\begin{equation*}
    |\phi_{n}\rangle_{\Sigma,{n+1}}^{\mathrm{symm}} 
    \equiv U_{n}(\phi_{n})|0\rangle_{n+1}|\phi_{n}\rangle_{\Sigma{n}}^{0},
\end{equation*}
where
\begin{equation*}
    |\phi_{n}\rangle_{\Sigma{n}}^{0} \equiv \sum_{p=0}^{\lfloor{n/2}\rfloor}a_{2p}^{0}(\phi_{n})|2p\rangle_{\Sigma{n}}.
\end{equation*}
We compare $|\phi_{n}\rangle_{\Sigma{n}}^{0}$
with $|\phi_{n-1}\rangle_{\Sigma{n}}^{\mathrm{symm}}$
for $n \ge 2$ to find an optimal angle.
Explicitly speaking, we compute the fidelity between the two states given as 
\begin{equation*}
    F_{n}(\phi_{n},\phi_{n-1}) \equiv \prescript{0}{\Sigma{n}}{\langle \phi_{n}|}\phi_{n-1}\rangle_{\Sigma{n}}^{\mathrm{symm}}
\end{equation*}
and set 
$\phi_{n-1}$ for a given $\phi_{n}$ by minimizing the infidelity as
\begin{equation}\label{eq:phi-n-minus-1-from-phi-n}
    \phi_{n-1} = \argmin_{\phi}[1 - F_{n}(\phi_{n},\phi)].
\end{equation}
The detailed derivation is given in Appendix~\ref{sec:N-M-ansatz}.
On the lower panel of Fig.~\ref{fig:atomic-ansatz},
we show the relation between $\phi_{n-1}$ and $\phi_{n}$ for several $n$.
We see that the bound given in Eq.~(\ref{eq:phi-n-range}) is preserved across layers.
Since $\phi_{n-1}$ is determined by $\phi_{n}$ for $2 \le n \le N-1$, 
all angles are parameterized by the last angle, $\phi_{N-1}$, for $N$ atomic qubits.
Setting $\phi_{N-1}=\phi$ for a given angle $0 \le \phi < \pi/2(N-1)$, we define the corresponding atomic state via Eq.~(\ref{eq:phi1-to-phi-Nminus1-ket-a}) as
\begin{equation}\label{eq:phi-ket-a}
    |\phi\rangle_{a} \equiv |{\phi_{1}(\phi_{2}),...,\phi_{N-2}(\phi),\phi}\rangle_{a}
\end{equation}
for a given $N>1$. For example, we have $|\phi\rangle_{a} = U_{1}(\phi)|0\rangle_{a}$ for $N=2$, and $|\phi\rangle_{a} = U_{2}(\phi)U_{1}(\phi_{1}(\phi))|0\rangle_{a} = |{\phi_{1}(\phi),\phi}\rangle_{a}$ for $N=3$ as given in Eq.~(\ref{eq:phi1-of-phi2-and-phi2-ket-a}). For $N=1$, we define $|\phi\rangle_{a} = |0\rangle_{a}$ as this is the only even-parity state and encompasses the $N=1$ ansatz given in Eq.~(\ref{eq:ansatz-N1-M}).
By using this $|\phi\rangle_{a}$ given in Eq.~(\ref{eq:phi-ket-a}) as the initial atomic state in Eq.~(\ref{eq:theta-phi-ket}), we finally obtain a general ansatz for given $N$ and $M$ as
\begin{equation}\label{eq:ansatz-N-M}
    |{\theta;\phi}\rangle \equiv e^{-i\theta X_{a}Y_{c}/2}|{\phi}\rangle_{a}|0\rangle_{c}.
\end{equation}
Below is the quantum circuit diagram for preparing the ansatz:
\begin{equation*}
    \Qcircuit @C=1em @R=.7em @!R { 
        & & & \dstick{M} & & \dstick{\theta} & \\
        & & \lstick{|0\rangle_{c}} & {/} \qw & \qw & \sgate{Y}{2} & \qw \\
        & & & \dstick{N} & & & \\
        & & \lstick{|0\rangle_{a}} & {/} \qw & \gate{U_{a}(\phi)} & \gate{X} & \qw\\
    }
\end{equation*}
where
\begin{equation*}
    U_{a}(\phi) \equiv U_{N-1}(\phi) U_{N-2}(\phi_{N-2}(\phi))\cdots U_{1}(\phi_{1}(\phi_{2}))
\end{equation*}
generates the atomic state $|\phi\rangle_{a}$ from $|0\rangle_{a}$.

\begin{figure}
    \centering
    \includegraphics[width=\linewidth]{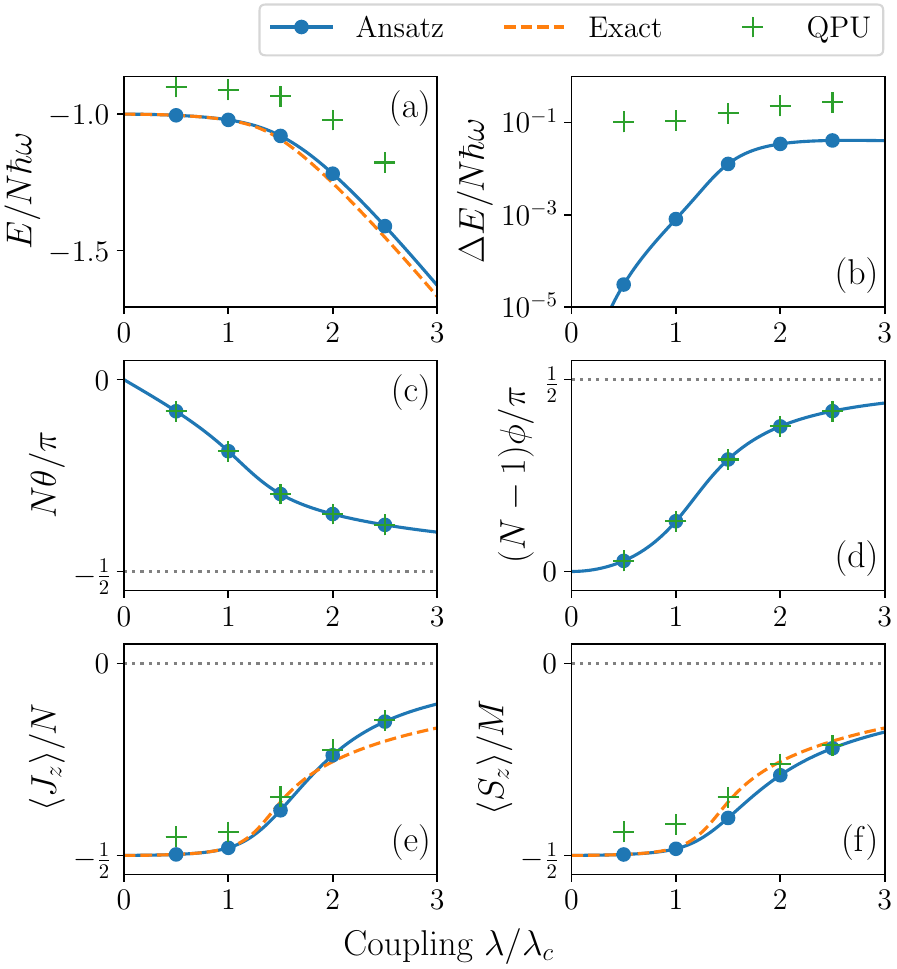}
    \caption{Energy optimization result with respect to the ansatz given in Eq.~(\ref{eq:ansatz-N-M}) for the spin-Dicke model with $N=8$ atomic qubits and $M=8$ cavity qubits. (a) Comparison of the optimized energy with respect to the ansatz (solid line with circles) and its implementation on a QPU (green crosses). The dashed line indicates the ground state energy $E_{G}$. (b) The difference $\Delta E = E - E_{G}$ between each energy $E$ shown in (a) and the ground state energy. (c) and (d) show optimized values of the ansatz angles, $\theta$ and $\phi$, respectively. The average atomic and cavity spins, $\langle J_{z}\rangle$ and $\langle S_{z}\rangle$, are given in (e) and (f), respectively. The gray dotted lines indicate strong-coupling asymptotic values.}
    \label{fig:ansatz-N-M}
\end{figure}

In Fig.~\ref{fig:ansatz-N-M}(a) and (b), we show optimization results of the energy expectation value with respect to the ansatz for the spin-Dicke Hamiltonian with $N=8$ atomic qubits and $M=8$ cavity qubits.
The angles $\theta$ and $\phi \equiv \phi_{N-1}$ optimizing the energy expectation value with respect to the ansatz are shown in panels (c) and (d). Their asymptotics toward the infinite coupling limit, $N\theta \rightarrow -\pi/2$ and $(N-1)\phi \rightarrow \pi/2$ as given respectively in Eqs.~(\ref{eq:theta-asympt-inf-coup}) and (\ref{eq:phi-n-inf-coup}), are indicated with green dotted lines.
We note that the ansatz follows qualitatively the ground state in terms of its energy as well as the order parameters as shown in panels (a), (e) and (f).
To test whether the ansatz can be implemented on a real device, 
we prepare the ansatz with the optimized values of the angles on QPU and measure the energy expectation value as well as the order parameters, indicated by the green crosses in panels (a), (e) and (f).
\subsection{Additional layer of ansatz for improving accuracy}
In this section, we present how to enhance the fidelity of the state prepared by the ansatz given in Eq.~(\ref{eq:ansatz-N-M}) with respect to the exact ground state.
We recall that we have used $\exp[-i\theta X_{a}Y_{c}/2]$ 
to entangle the atomic and the cavity qubits in the ansatz $|{\theta;\phi}\rangle$.
However, this was not the only operation which preserves the parity, time-reversal and exchange symmetries; we could have selected $\exp[-i\theta Y_{a}X_{c}/2]$.
The reason we have chosen the former, generated by $X_{a}Y_{c}$, is that this gives the correct asymptotic limit of the constructed ansatz as shown in Sec.~\ref{sec:ansatz-N1-M} and Sec.~\ref{sec:ansatz-N-M}.
However, if we use both operators, the former can satisfy the asymptotic condition while the latter can correct residual error in the phase transition region.
Thus, on top of the ansatz
given in Eq.~(\ref{eq:ansatz-N-M}),
we propose 
\begin{equation}\label{eq:ansatz-N-M-L2}
    |{\theta_{1},\theta_{2};\phi}\rangle = e^{-i\theta_{2}Y_{a}X_{c}/2}|{\theta_{1};\phi}\rangle,
\end{equation}
where we add the collective MS gates generated by $Y_{a}X_{c}$.
Below, we show the quantum circuit for preparing the ansatz:
\begin{equation*}
    \Qcircuit @C=1em @R=.7em @!R { 
        & & & \dstick{M} & & \dstick{\theta_{1}} & \dstick{\theta_{2}} & \\
        & & \lstick{|0\rangle_{c}} & {/} \qw & \qw & \sgate{Y}{2} & \sgate{X}{2} & \qw\\
        & & & \dstick{N} & & & &\\
        & & \lstick{|0\rangle_{a}} & {/} \qw & \gate{U_{a}(\phi)} & \gate{X} & \gate{Y} & \qw\\
        \\
    }
\end{equation*}
We note that this second layer of atomic-cavity entangling gates, parameterized by $\theta_{2}$, does not commute with the first layer in general, except for trivial angles or for $N=M=1$.
We expect $\theta_{2}$ to be small compared to the angle of the first layer, $\theta_{1}$, since the first layer already reduces the relative energy deviation $\Delta{E}/|E_{G}|$ to few percent as can be seen by the blue solid lines in Fig.~\ref{fig:ansatz-N-M}(a) and (b) for $N=M=8$.

\begin{figure}
    \centering
    \includegraphics[width=\linewidth]{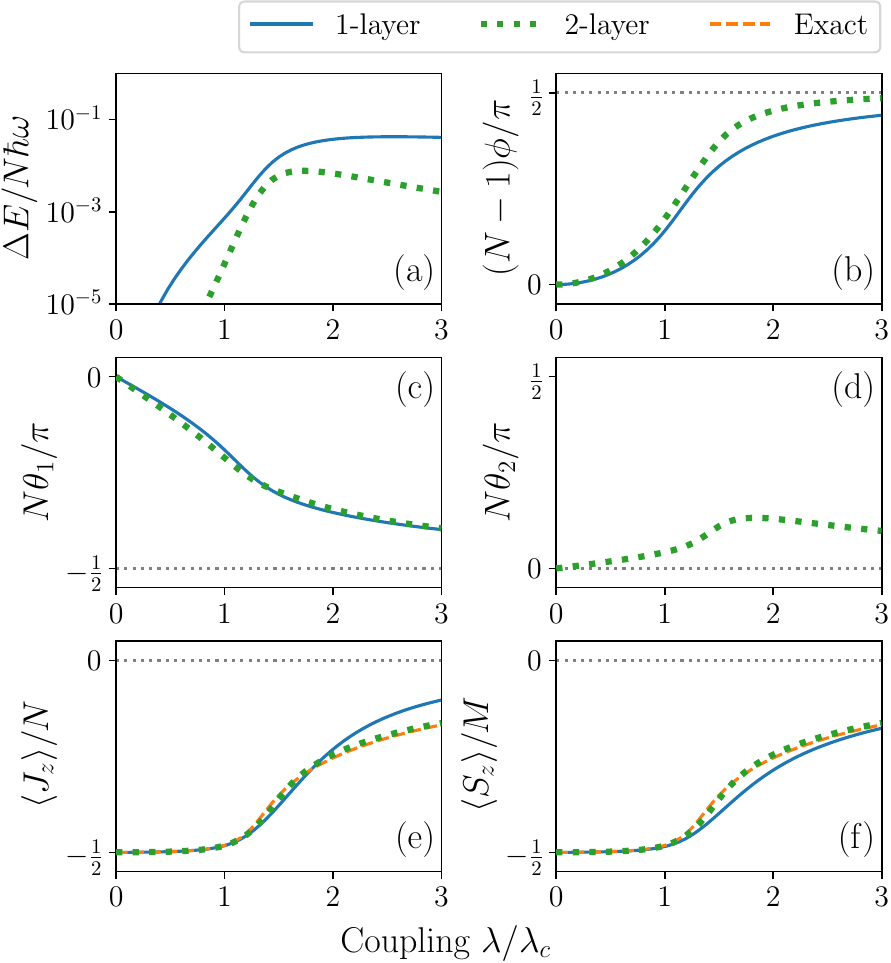}
    \caption{Comparison of the one- and two-layer ans\"{a}tze given in Eqs.~(\ref{eq:ansatz-N-M}) and (\ref{eq:ansatz-N-M-L2}), respectively, for the spin-Dicke model with $N=M=10$. (a) Difference $\Delta E = \langle H\rangle - E_{G}$ between the optimized expectation value of the Hamiltonian with respect to each ansatz and the ground-state energy. (b,c,d) Optimized values of angles for each ansatz. (e,f) Expectation values of collective spins, $J_{z}$ and $S_{z}$, per particle. Gray dotted lines indicate the infinite-coupling asymptotes.}
    \label{fig:one-and-two-layer-ansatze}
\end{figure}

In Fig.~\ref{fig:one-and-two-layer-ansatze}, we show optimized energy expectation value with respect to the one- and two-layer ans\"{a}tze for the spin-Dicke model with $N=M=10$. By adding the second layer, the energy difference $\Delta E = \langle H\rangle - E_{G}$ between the ansatz and the true ground state decreases as shown in panel (a). The smallest enhancement (i.e., reduction) in $\Delta E$ occurs within $\lambda_{c} < \lambda <1.5\lambda_{c}$, with the ratio between the one-layer $\Delta E$ with respect to that of the two-layer ansatz is around $2$.
Comparing panels (c) and (d),
we see that the second angle $\theta_{2}$ is small compared to the first in magnitude, correcting the order parameters shown in panels (e) and (f), which can allow better estimation of the phase transition point once the physical device can execute the quantum circuit with a sufficiently low error rate.



\section{Dicke model on hybrid qubit-boson simulators}\label{sec:ansatz-spin-boson-hybrid-for-dicke}

In this section, we present an ansatz to probe the ground state of the Dicke model on a hybrid qubit-boson quantum simulator, based on the results on the spin-Dicke model. The bosonic aspect of the computation can be achieved by the collective phonon modes encountered in trapped ion hardware or the microwave resonator in superconducting computing. The remaining fermionic operations are achieved by ''standard" single-qubit gates.
 
If we use a sufficiently large $M$ for a given $N$ and $\lambda$,
the ground state of the Dicke model can be approximated with the spin-Dicke model with high fidelity. 
However, it is also possible to use a physical bosonic mode directly, without mapping it onto a set of physical qubits.

In this case, the atom-cavity entangling operation in the Spin-Dicke model can be mapped back to 
\begin{equation*}
    e^{-i\theta X_{a}Y_{c}/2} 
    \mapsto 
    e^{-ix X_{a}(ia^{\dagger}-ia)/\sqrt{2}},
\end{equation*}
where we use $Y_{c} = iS_{+} - iS_{-} \mapsto \sqrt{M}(ia^{\dagger}-ia)$, cf. Eq.~(\ref{eq:a-to-Sminus}), and normalize the parameter as 
$\theta\sqrt{M/2} \equiv x$.
Introducing the $P$ quadrature operator of the bosonic mode as $P \equiv (ia^{\dagger}-ia)/\sqrt{2}$, the ansatz for a finite $M$ transforms to
\begin{equation}\label{eq:ansatz-N}
    |{x;\phi}\rangle 
    = e^{-ix X_{a}P}|\phi\rangle_{a}|0\rangle_{c},
\end{equation}
where the atomic state is $|\phi\rangle_{a}$ prepared by Eq.~(\ref{eq:phi1-to-phi-Nminus1-ket-a}) for a given $\phi = \phi_{N-1}$ and $\phi_{n-1}$ as a function of $\phi_{n}$.
This spin-boson hybrid operation can be interpreted as spin-dependent displacement $D[xX_{a}/\sqrt{2}]$ or boson-dependent spin rotation $R_{X_{a}}[2xP]$.
Under this entangling operation, 
the Dicke Hamiltonian transforms via
\begin{equation*}
    Z_{a} 
    \rightarrow \cos(2xP)Z_{a} + \sin(2xP)Y_{a},
\end{equation*}
and
\begin{equation*}
    a 
    \rightarrow a + x X_{a}/\sqrt{2}.
\end{equation*}
Taking the expectation value of the transformed Hamiltonian with respect to the product state of $|\phi\rangle_{a}$ and the bosonic vacuum state $|0\rangle_{c}$, we have
\begin{equation*}
\begin{split}
    \langle H \rangle(x;\phi) 
    &= \left(\omega \frac{x^{2}}{2} + \frac{\lambda}{\sqrt{N}}\sqrt{2}x\right){}_{a}\langle \phi | X_{a}^{2}|\phi \rangle_{a}\\
    & - \frac{\omega_{0}}{2} {}_{a}\langle \phi | Z_{a}|\phi \rangle_{a} \times {}_{c}\langle 0 | \cos(2xP) |0 \rangle_{c}.
\end{split}
\end{equation*}
Evaluating the expectation value on the vacuum state as
\begin{equation*}
    {}_{c}\langle 0 | \cos(2xP) |0 \rangle_{c} 
    = \frac{1}{2}{}_{c}\langle 0 | D[\sqrt{2}x] |0 \rangle_{c} + \mathrm{c.c.} = e^{-x^{2}},
\end{equation*}
we have
\begin{equation}\label{eq:dicke-H-avg}
    \langle H \rangle(x;\phi)
    = x\left(\frac{\omega}{2}x + \frac{\sqrt{2}\lambda}{\sqrt{N}}\right)\langle X_{a}^{2} \rangle_{a}(\phi)
    - \frac{\omega_{0}}{2} e^{-x^{2}}\langle Z_{a}\rangle_{a}(\phi)
\end{equation}
for $\langle O\rangle_{a}(\phi) \equiv {}_{a}\langle \phi | O|\phi \rangle_{a}$.

For $N=1$, 
the initial atomic state is $|\phi\rangle_{a} = |0\rangle_{a}$ since this is the only even-parity state of a single atomic qubit. Thus, the ansatz for $N=1$ does not depend on $\phi$ and the corresponding energy expectation values reads
\begin{equation*}
    \langle H \rangle(x;\phi)
    = x\left(\frac{\omega}{2}x + \sqrt{2}{\lambda}\right)
    - \frac{\omega_{0}}{2} e^{-x^{2}}.
\end{equation*}
For $N \ge 2$, the energy expectation value depends also on $\phi$.
Taking $N=2$ for instance, computing 
$\langle Z_{a}\rangle_{a}(\phi) 
= 2\cos{\phi}$ 
and $\langle X_{a}^{2}\rangle_{a}(\phi) 
= 2(1 + \sin{\phi})$ gives
\begin{equation*}
    \langle H \rangle(x;\phi)
    = 2x\left(\frac{\omega}{2}x + \lambda\right)(1 + \sin{\phi})
    - \omega_{0} e^{-x^{2}}\cos{\phi}.
\end{equation*}
For general $N$, $\langle X_{a}^{2}\rangle_{a}(\phi)$ and $\langle Z_{a}\rangle_{a}(\phi)$ are computed 
by constructing $|\phi\rangle_{a}$ via Eq.~(\ref{eq:phi1-to-phi-Nminus1-ket-a}) with $\phi_{N-1}=\phi$ and the rest of the angles, $\phi_{n}$ for $1 \le n \le N-2$, determined by Eq.~(\ref{eq:phi-n-minus-1-from-phi-n}).

In the strong-coupling limit, the energy expectation value is dominated by the first term in Eq.~(\ref{eq:dicke-H-avg}) as
\begin{equation*}
    \langle H \rangle(x;\phi)
    \xrightarrow[]{\lambda \gg \lambda_{c}} x\left(\frac{\omega}{2}x + \frac{\sqrt{2}\lambda}{\sqrt{N}}\right)\langle X_{a}^{2} \rangle_{a}(\phi)
\end{equation*}
since the second term is lower-bounded by $-N\omega_{0}/2$. 
Minimizing the quadratic polynomial of $x$ by making a square, 
and maximizing $\langle X_{a}^{2} \rangle_{a}(\phi) \le N^{2}$ 
by setting $|\phi\rangle_{a}$ as 
Eq.~(\ref{eq:phi-ket-a-inf-coup}), the energy expectation value in the limit is minimized as
\begin{equation}\label{eq:H-avg-x-phi-strong-coup-regime}
    \min_{x,\phi}
    \frac{\langle H \rangle(x;\phi)}{N\omega}
    \xrightarrow[]{\lambda \gg \lambda_{c}} 
    - \left(\frac{\lambda}{\omega}\right)^{2}
\end{equation}
via 
\begin{equation}\label{eq:x-asymptote-strong-coup}
\begin{split}
    x 
    &\rightarrow -\frac{\sqrt{2}}{\sqrt{N}}\frac{\lambda}{\omega}
\end{split}
\end{equation}
as well as $\phi \rightarrow {\pi}/{2}(N-1)$ following from Eq.~(\ref{eq:phi-n-inf-coup}).
The corresponding asymptotic limit of the average boson number per atom is given as $\langle a^{\dagger}a\rangle/N \rightarrow (\lambda/\omega)^{2}$.

\begin{figure}
    \centering
    \includegraphics[width=\linewidth]{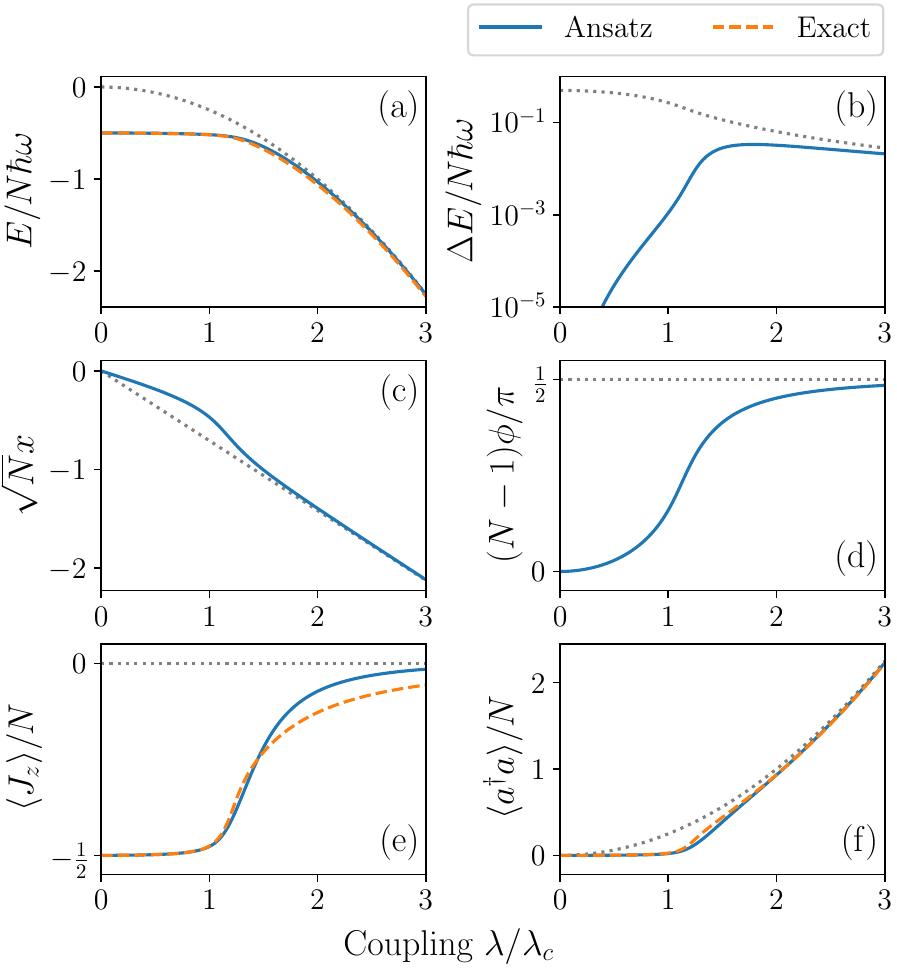}
    \caption{
        (a) Ground-state energy of the Dicke model describing $N=10$ atoms resonant to the cavity mode (orange dashed line). The blue solid line indicates the optimized energy expectation value with respect to the ansatz, Eq.~(\ref{eq:ansatz-N}). The strong-coupling asymptote given in Eq.~(\ref{eq:H-avg-x-phi-strong-coup-regime}) is shown with the gray dotted line. (b) Difference in energy, $\Delta E = \langle H\rangle - E_{G}$, between the ansatz and the ground state. The gray dotted line indicates the difference between the energy asymptote and the ground-state energy. (c,d) Optimal values of the ansatz parameters: spin-dependent displacement amplitude $x$ and the rotational angle $\phi$ for preparing the initial atomic state. The asymptotes of $x$ and $\phi \equiv \phi_{N-1}$ are given in Eqs.~(\ref{eq:x-asymptote-strong-coup}) and (\ref{eq:phi-n-inf-coup}), respectively. (e,f) Order parameters of the ground state and the optimized ansatz. Strong-coupling asymptotes are indicated by the gray dotted lines.
    }
    \label{fig:N}
\end{figure}
In Fig.~\ref{fig:N}, we show optimization result of the proposed ansatz for the Dicke model with $N=10$ particles at resonance to the cavity mode. The difference shown in Fig.~\ref{fig:N}(b), $\Delta E = \langle H \rangle - E_{G}$, between the ground-state energy and the expectation value optimized with respect to the ansatz is less than 4\% of $N\hbar\omega$. As indicated by the dashed gray line in each panel, the phase boundary is visible at $\lambda = \lambda_{c}$, in terms of energy and order parameters shown in Fig.~\ref{fig:N}(a), (e) and (f).

To realize the Dicke ansatz given in Eq.~(\ref{eq:ansatz-N}), we need physical implementation of the qubit-qubit entangling as well as the qubit-bosonic hybrid operations. 
In particular, bosonic displacement conditional on a qubit state is essential. 
This operation is a building block of universal qubit-oscillator hybrid quantum computation~\cite{Eickbusch2022} and has been demonstrated on systems 
including trapped ions~\cite{Leibfried2003,PhysRevLett.116.140402} and superconducting circuits~\cite{Campagne-Ibarcq2020}.
We see that there is a limitation regarding the maximum available average boson number which can be implemented on near-term quantum devices. In trapped-ion processors, high-fidelity operations rely on the Lamb-Dicke approximation, which restricts the average phonon number around 10 to 20 in current experiments~\cite{cai2021observation,PhysRevLett.116.140402,PhysRevX.8.021027}. In superconducting circuit-QED architectures, oscillator occupation is instead limited by the critical photon number associated with the dispersive approximation, whose value depends on the qubit–resonator coupling strength and detuning~\cite{Eickbusch2022,RevModPhys.93.025005}. 
However, since the phase transition is of second order, the average boson number can be small near the critical point; the discontinuity in the thermodynamic limit occurs in the \textit{derivative} of the average boson number, not in the average boson number itself, so that it can be small near the critical point. Thus, to observe quantum phase transition on such quantum processing units with qubit-oscillator hybrid degrees of freedom, it may be sufficient to investigate the behavior of the system near the critical point, with a less demanding requirement on the maximum boson number that can be measured.

\section{Conclusions and Discussion}

In this work, we developed a variational quantum framework for studying the finite-size Dicke model on both fully digital and hybrid quantum computing platforms. For digital quantum processors, we introduced a spin-Dicke representation based on the inverse Holstein-Primakoff transformation, in which the bosonic cavity mode is encoded in a finite register of qubits. By linearizing the inverse transformation, this mapping yields an effective finite-dimensional spin-Dicke model that provides a systematically improvable approximation to the Dicke Hamiltonian as the cavity truncation size is increased, while remaining suitable for implementation on the near-term digital quantum processors. Building upon this representation, we developed symmetry-preserving variational ansätze that explicitly exploit the parity, time-reversal, and exchange symmetries of the finite-size Hamiltonian. Full VQE simulations on an ideal statevector simulator demonstrate that these ansätze accurately reproduce both the ground- and excited-state properties across the parameter regimes considered, exhibiting excellent agreement with exact diagonalization. Beyond benchmarking, we performed a systematic analysis of the variational manifold by examining the analytical structure of the ansätze and the behavior of the optimized variational parameters. This analysis reveals how the underlying symmetries constrain the optimization landscape, identifies redundant variational degrees of freedom, and provides general design principles for constructing scalable symmetry-preserving ansätze.

To assess the experimental feasibility of the proposed circuits, we prepared the classically optimized variational states on a trapped-ion quantum processor. Rather than performing a hardware-in-the-loop VQE optimization, the optimized parameters obtained from the classical optimization were used directly for state preparation and measurement of the relevant observables. This proof-of-principle experiment demonstrates that the proposed symmetry-preserving circuits can be implemented on current trapped-ion hardware.

Finally, we introduced a complementary hybrid qubit--bosonic variational ansatz that directly exploits bosonic degrees of freedom, thereby reducing both the required number of qubits and the circuit depth compared with fully digital implementations. Together, the digital and hybrid approaches presented here establish a symmetry-aware framework for variational quantum simulations of finite-size light--matter systems.  Future work will focus on extending these methods to larger cavity truncations, incorporating hardware-aware compilation and error-mitigation techniques, and applying them to multimode cavity-QED systems, inhomogeneous Dicke models, and other strongly correlated spin--boson Hamiltonians.

\subsection*{Acknowledgments}
This project has been supported by the National Research Foundation of the Republic of Korea (NRF) (Grant No. RS-2024-00509579) and the Luxembourg National Research Fund (FNR) (project INTER/Korea/24/19350574) through the Luxembourg-Korea joint research pilot programme. S.A., A.G.L. and A.S.M. acknowledge also support by the NRF also via Grant No. RS-2026-25483912.
Computing time on QPU has been acquired by the Quantum Information Science R\&D Ecosystem Creation program funded by NRF (Grant No. 2020M3H3A1110365) and the IonQ research credits program.
A.~P.~B. acknowledges the financial support from the Department of Science and Technology (DST), Government of India, through the INSPIRE Faculty Fellowship (Ref no. IFA24-PH 314).
S.~A. has been supported by the Jang Young Sil Fellow Program of KAIST, Republic of Korea.

\subsection*{Code availability}

Code is available upon reasonable requests.

\appendix

\section{Resource estimates and limits for the classical simulation of Dicke model}\label{sec:classical_simulability}
For a generalized spin-$j$ system coupled to a spin-$s$ particle, we can explicitly write the Hamiltonian as

\begin{align*} 
    H &= \omega\left[I^{\otimes N}\otimes S_{z}\right] + \sum_{n=1}^{N} \omega_{n} \left[I^{\otimes n-1} \otimes J_{z} \otimes I^{\otimes N+1-n} \right]\\
    &+ \sum_{n=1}^{N}\frac{2^{2}\lambda_{n}}{\sqrt{N}} \left [I^{\otimes n-1} \otimes J_{x} \otimes I^{\otimes N-n} \otimes S_{x}\right], 
\end{align*}
The identity matrices for spin-$j$ and spin-$s$ system have dimensions $2j+1$ and $2s+1$, respectively, resulting in Hamiltonian matrix dimension of $\left[(2s+1)\cdot (2j+1)^{N-1}\right]^2$. If we consider a simpler spin-$1/2$ system, the $J$, $S$ matrices are replaced by their corresponding Pauli operators, resulting in a Hamiltonian of dimension $2^{2N}$. The exponential growth in Hilbert space dimension makes storing these states and operators rapidly untenable on classical hardware. The default datatype used for most matrix computations is float64, which requires 8 bytes for each number, even assuming access to one of the worlds most powerful supercomputers "Frontier", which has $\sim 512\cdot9856$ gigabytes of RAM storage~\cite{FrontierGuide}, the largest number of qubits that could be considered is 52.

More realistic setups and experiments are likely to correspond to the non-homogeneous case where $\omega_{n}$ and $\lambda_{n}$ values differ. In these situations our system cannot be analytically solved since not all coupling strengths and frequencies will be equal, thus classical simulation offers a useful starting point for probing potential interesting physical behavior. The presence of tensor products with identity matrices can be exploited via the resulting sparse nature of the Hamiltonian. One can avoid the explicit construction of the Hamiltonian by considering only its action on a vector instead and to use corresponding tools \cite{SciPyEigshDoc} to find the ground and excited state energies with their corresponding eigenstates.

\section{Holstein-Primakoff transformation}\label{sec:HPT}
Consider a subspace $\{|0\rangle,\cdots,|2s\rangle\}$ in a Fock space for an integer $2s\ge 1$.
Mapping $|n\rangle$ to an eigenstate $|m={-s+n}\rangle$ of $S_{z}$,
the number operator in the subspace takes the form

\begin{equation*}
    a^{\dagger}a 
    = \sum_{n=0}^{2s}|n\rangle n \langle n|
    \mapsto \sum_{m=-s}^{s} |m\rangle (s+m) \langle m| 
    = s + S_{z}.
\end{equation*}
Likewise, the annihilation operator can be mapped as
\begin{equation*}
\begin{split}
    a 
    &= \sum_{n=0}^{2s-1}|n\rangle \sqrt{n+1} \,\langle n+1|
    \mapsto \sum_{m=-s}^{s-1}|m\rangle \sqrt{m+s+1}\,\langle m+1|.
\end{split}
\end{equation*}
Using $\langle m|S_{-} = \langle m+1|\sqrt{2s-(m+s)}\sqrt{m+s+1}$, we get
\begin{equation*}
    a \mapsto \sum_{m=-s}^{s} |m\rangle\frac{1}{\sqrt{2s-(m+s)}}\langle m|S_{-}
    = \frac{1}{\sqrt{2s - (S_{z}+s)}}S_{-},
\end{equation*}
where the upper limit of the summation could be changed from $s-1$ to $s$ due to $\langle s|S_{-} = 0$.

\section{Single-layer energy expectation value}\label{sec:single-layer-energy-expectation-value}
In this section, we derive the energy expectation value, Eq.~(\ref{eq:avg-H-ansatz-N1-M}).
The ansatz is
\begin{equation*}
    |\theta \rangle = e^{-i\theta X_{a}Y_{c}/2}|0\rangle,
\end{equation*}
where the zero-angle state is the zero-excitation state as $|0\rangle = |0\rangle_{a}|0\rangle_{c}$.
Then, the expectation value reads
\begin{equation*}
    \langle \theta | H |\theta \rangle = \langle 0 | e^{i\theta X_{a}Y_{c}/2} H e^{-i\theta X_{a}Y_{c}/2}|0 \rangle.
\end{equation*}
We can view the collective gate as a rotational operator in two different axes,
\begin{equation*}
    e^{-i\theta X_{a}Y_{c}/2} = R_{X_{a}}(\theta Y_{c}) = R_{Y_{c}}(\theta X_{a}),
\end{equation*}
where $R_{A}(\theta) \equiv \exp[-i\theta A/2]$.
Transforming each term of the Hamiltonian with respect to the collective gate via
\begin{equation*}
\begin{split}
     R_{Y_{c}}^{\dagger}(\theta X_{a}) Z_{c} R_{Y_{c}}(\theta X_{a})
     &= \cos{(\theta X_{a})} Z_{c} - \sin{(\theta X_{a})} X_{c},\\
     R_{X_{a}}^{\dagger}(\theta Y_{c}) Z_{a} R_{X_{a}}(\theta Y_{c})
     &= \cos{(\theta Y_{c})} Z_{a} - \sin{(\theta Y_{c})} Y_{a},\\
     R_{Y_{c}}^{\dagger}(\theta X_{a}) X_{c}R_{Y_{c}}(\theta X_{a})
     &= \cos{(\theta X_{a})} X_{c} + \sin{(\theta X_{a})} Z_{c},
\end{split}
\end{equation*}
and noting that $|0\rangle$ is the product of $|0\rangle_{a}$ and $|0\rangle_{c}$,
we can compute each contribution as
\begin{equation}\label{eq:avg-Zc-Za-XaXc-R_XaYc}
\begin{split}
    \langle \theta | Z_{c} | \theta \rangle / M &= {}_{a}\langle 0|\cos{(\theta X_{a})}|0\rangle_{a},\\
    \langle \theta | Z_{a} | \theta \rangle / N &= {}_{c}\langle 0|\cos{(\theta Y_{c})}|0\rangle_{c},\\
    \langle \theta | X_{a}X_{c} | \theta \rangle / M &= {}_{a}\langle 0|X_{a}\sin{(\theta X_{a})}|0\rangle_{a},    
\end{split}
\end{equation}
since ${}_{c}\langle 0|X_{c}|0\rangle_{c} = 0$, ${}_{a}\langle 0|Y_{a}|0\rangle_{a}=0$, ${}_{c}\langle 0|Z_{c}|0\rangle_{c} = M$ and ${}_{a}\langle 0|Z_{a}|0\rangle_{a} = N$.
Writing the cosine function as
\begin{equation*}
    {}_{a}\langle 0|\cos{(\theta X_{a})}|0\rangle_{a} = \frac{1}{2}{}_{a}\langle 0|e^{i\theta X_{a}}|0\rangle_{a} + \mathrm{c.c.},
\end{equation*}
and decomposing the exponential into individual qubit subspaces,
\begin{equation*}
    {}_{a}\langle 0|e^{i\theta X_{a}}|0\rangle_{a} = \prod_{n=1}^{N}{}_{an}\langle 0|e^{i\theta X_{an}}|0\rangle_{an}
    = \cos^{N}(\theta),
\end{equation*}
we get ${}_{a}\langle 0|\cos{(\theta X_{a})}|0\rangle_{a}$ and, similarly, ${}_{c}\langle 0|\cos{(\theta Y_{c})}|0\rangle_{c}$.
The derivative of the expectation value with respect to the angle $\theta$ gives 
\begin{equation*}
    {}_{a}\langle 0|X_{a}\sin{(\theta X_{a})}|0\rangle_{a} = -\partial_{\theta}{}_{a}\langle 0|\cos{(\theta X_{a})}|0\rangle_{a}.
\end{equation*}
Thus, 
Eq.~(\ref{eq:avg-Zc-Za-XaXc-R_XaYc}) becomes
\begin{equation}\label{eq:avg-theta-ket}
\begin{split}
    \langle \theta | Z_{c} | \theta \rangle &= M \cos^{N}(\theta),\\
    \langle \theta | Z_{a} | \theta \rangle &= N \cos^{M}(\theta),\\
    \langle \theta | X_{a}X_{c} | \theta \rangle &= MN\cos^{N-1}(\theta)\sin{(\theta)},
\end{split}
\end{equation}
for which
\begin{equation*}
\begin{split}
    \langle \theta | H |\theta \rangle
    &= 
    -\frac{\omega}{2}M \cos^{N}(\theta)
    -\frac{\omega_{0}}{2}N \cos^{M}(\theta)\\
    &+ {\lambda}\sqrt{MN}\cos^{N-1}(\theta)\sin{(\theta)}
\end{split}
\end{equation*}
Setting $N=1$ gives Eq.~(\ref{eq:avg-H-ansatz-N1-M}).

\section{Implementation on trapped ions}\label{sec:qpu}


For realizing the proposed circuit on a quantum device, we use $ZZ$ gate, $R_{Z_{j}Z_{k}}(\phi) = \exp[-i\phi Z_{j}Z_{k}/2]$, and single-qubit $\pi/2$ rotations as native gates. For QPU backends, we utilized IonQ Forte for Fig.~\ref{fig:spin-dicke-N1-multiM} and Forte Enterprise for Fig.~\ref{fig:ansatz-N-M}.
Since the ans\"{a}tze, Eqs.~(\ref{eq:ansatz-N1-M}) and (\ref{eq:ansatz-N-M}), consist of $XY$ gates, we convert each of them into $ZZ$ gate and single-qubit rotations so that the circuit are composed only of the native gates.
As an example, we show below the quantum circuit for generating the spin-Dicke ansatz for $N=4$ and $M\ge1$, denoting the single-qubit $\pi/2$ rotation generated by a Pauli operator $A \in \{X,-X,Y,-Y\}$ as $R_{A} \equiv \exp[-i\pi A/4]$,
\begin{widetext}
\begin{equation*}
    \Qcircuit @C=1em @R=.8em @!R { 
        & & \dstick{M} & 
        & & & & & & & & &
        & & & & & &
        \\
        & \lstick{c\,|0\rangle} & {/} \qw & \gate{R_{-X}} 
        & \qw & \qw & \qw & \qw & \qw & \qw & \qw & \qw & \qw
        & \sgate{Z}{1} & \sgate{Z}{2} & \sgate{Z}{3} & \sgate{Z}{4} & \gate{R_{X}} & \meter
        \\
        & \lstick{a1\,|0\rangle} & \qw & \gate{R_{-Y}} 
        & \sgate{Z}{1} & \qw & \sgate{Z}{2} & \qw & \qw & \sgate{Z}{3} & \qw & \qw & \qw
        & \gate{Z} & \qw & \qw & \qw & \gate{R_{Y}} & \meter
        \\
        & \lstick{a2\,|0\rangle} & \qw & \gate{R_{-Y}} 
        & \gate{Z} \gategroup{3}{5}{4}{5}{.7em}{--} & \gate{R_{-X}} & \qw & \sgate{Z}{1} & \qw & \qw & \sgate{Z}{2} & \qw & \qw
        & \qw & \gate{Z} & \qw & \qw & \gate{R_{Y}} & \meter
        \\
        & \lstick{a3\,|0\rangle} & \qw & \gate{R_{-Y}} 
        & \qw & \qw & \gate{Z} & \gate{Z} \gategroup{3}{7}{5}{8}{.7em}{--} & \gate{R_{-X}} & \qw & \qw & \sgate{Z}{1} & \qw
        & \qw & \qw & \gate{Z} & \qw & \gate{R_{Y}} & \meter
        \\
        & \lstick{a4\,|0\rangle} & \qw & \gate{R_{-Y}} 
        & \qw & \qw & \qw & \qw & \qw & \gate{Z} & \gate{Z} & \gate{Z} \gategroup{3}{10}{6}{12}{.7em}{--} & \gate{R_{-X}}
        & \qw & \qw & \qw & \gate{Z} \gategroup{2}{14}{6}{17}{.7em}{.} & \gate{R_{Y}} & \meter
        \\
        \\
    }
\end{equation*}
\end{widetext}
On the right side, there are $ZZ$ gates enclosed by a dotted line. Each gate is parameterized by $-\theta$, where the angle $\theta \le 0$ is identical to the ansatz parameter in Eq.~(\ref{eq:ansatz-N-M}).
In addition, there are three groups with dashed boxes consisting of one, two, and three $ZZ$ gates. They are parameterized by $\phi_{1}$, $\phi_{2}$, and $\phi_{3}$, respectively, which are the angles introduced in Eq.~(\ref{eq:phi1-to-phi-Nminus1-ket-a}) and interrelated by Eq.~(\ref{eq:phi-n-minus-1-from-phi-n}).

We repeat preparing the ansatz and measure the qubits in the computational basis to estimate
$\langle Z_{a}\rangle$ and $\langle Z_{c}\rangle$ in the Hamiltonian, Eq.~(\ref{eq:H-qubits}). To estimate the remaining term in the Hamiltonian, $\langle X_{a}X_{c}\rangle$, we rotate the measurement axis by appending $R_{-Y}$ gate for each of the qubit at the end of the circuit and measure in the computational basis. 
We note that the $R_{-Y}$ gates appended for the atomic qubits cancel out with the $R_{Y}$ gate already presented in the circuit above. For the cavity qubits, each $R_{X}$ gate at the rightmost position in the circuit merges with the appended $R_{-Y}$ to create $R_{-Y}R_{X}$, which is equivalent to $R_{-Y}R_{X}R_{Y}R_{-Y} = R_{Z}R_{-Y}$, and the last $R_{Z}$ merges to the meter as it is diagonal in the computational basis.
We use $1000$ samples for preparation of the $N=1$ \& $M=35$ ansatz shown in Fig.~\ref{fig:spin-dicke-N1-multiM} and $4000$ samples for the $N=M=8$ ansatz presented in Fig.~\ref{fig:ansatz-N-M}.

We note that it is possible to execute multiple two-qubit gates simultaneously for a trapped-ion platform using global addressing of laser on the corresponding ions and nonlocal interaction via vibrational modes of the ionic crystal~\cite{Lu2019Global}. This gives possibility of compressing the circuit to have linear depth in the number of particles, $N$ and $M$. To realize this scheme, the laser parameters including the power, frequency, and phase, on each ion shall be tuned to implement the correct connectivity, e.g., one-to-all as given in the circuit, and the angle of each two-qubit gate. Finding the correct parameters is a classical optimization task, since the operation remains in the Lamb-Dicke limit and we need only one Pauli operator per ion for which the Magnus expansion terminates at the second order under rotating wave approximation~\cite{PhysRevLett.97.050505}. If the laser power is limited to address all ions at once, it is also possible to group two-qubit gates into a collection of multi-qubit gates which compresses the circuit, e.g., by a factor of two if we execute each pair of $ZZ$ gates simultaneously.

\section{Entangling atomic qubits for spin-Dicke ansatz}\label{sec:N-M-ansatz}

We use a series of the following operations composed of the $XY$-type MS gates to construct an ansatz with correct symmetries up to $N=4$:
\begin{equation*}
    U_{n}(\phi) \equiv \exp\left[-i \frac{\phi}{2}Y_{n+1}X_{\Sigma n}\right],
\end{equation*}
where $X_{\Sigma n} = \sum_{m=1}^{n}X_{m}$. Note that $U_{n}$ is symmetric under any exchange between qubits with indices from $1$ to $n$, but in general not with respect to such an operation involving the qubit with index $n+1$.
We sequentially apply the gate to the initial 
state of the qubits, $|0\rangle_{\Sigma N} = |0\rangle_{N}\cdots|0\rangle_{1}$, 
as
\begin{equation}\label{eq:even-num-init-state-generation-subroutine}
    \prod_{n=1}^{N-1}U_{n}(\phi_{n})|0\rangle_{\Sigma N}.
\end{equation}
Let us denote
a state of the first $n$ qubits as $|\cdot\rangle_{\Sigma{n}}$. 
The first operation of the subroutine, Eq.~(\ref{eq:even-num-init-state-generation-subroutine}), acts on the first two qubits in their ground state, $|0\rangle_{\Sigma{2}} = |0\rangle_{2}|0\rangle_{\Sigma{1}}=|0\rangle_{2}|0\rangle_{1}$, as
\begin{equation*}
\begin{split}
    |\phi_{1}\rangle_{\Sigma{2}} 
    &\equiv U_{1}(\phi_{1})|0\rangle_{\Sigma{2}} \\
    &= \cos\left(\frac{\phi_{1}}{2}\right)
    |0\rangle_{2}|0\rangle_{\Sigma{1}}
    + \sin\left(\frac{\phi_{1}}{2}\right)
    |1\rangle_{2}|1\rangle_{\Sigma{1}}.
\end{split}
\end{equation*}
For $N=3$ atomic qubits, we apply the next layer to link the third qubit with the first two qubits as
\begin{equation}\label{eq:phi1-phi2-ket-Sigma3}
\begin{split}
    &|\phi_{1},\phi_{2}\rangle_{\Sigma{3}}
    \equiv U_{2}(\phi_{2})|0\rangle_{3}|\phi_{1}\rangle_{\Sigma{2}}\\
    &= f_{0,0}|0\rangle_{3}|0\rangle_{\Sigma{2}}
    + f_{1,1}
    |1\rangle_{3}|1\rangle_{\Sigma{2}}
    + f_{0,2}
    |0\rangle_{3}|2\rangle_{\Sigma{2}},
\end{split}
\end{equation}
where
\begin{equation*}
\begin{split}
    f_{0,0}(\phi_{1},\phi_{2}) 
    &= \cos\left(\frac{\phi_{1}}{2}\right)\cos^{2}\left(\frac{\phi_{2}}{2}\right)
    - \sin\left(\frac{\phi_{1}}{2}\right)\sin^{2}\left(\frac{\phi_{2}}{2}\right),
    \\
    f_{1,1}(\phi_{1},\phi_{2})
    &= 
    \sqrt{2} \left[\cos\left(\frac{\phi_{1}}{2}\right) + \sin\left(\frac{\phi_{1}}{2}\right)\right]
    \cos\left(\frac{\phi_{2}}{2}\right)\sin\left(\frac{\phi_{2}}{2}\right),
    \\
    f_{0,2}(\phi_{1},\phi_{2})
    &= 
    \sin\left(\frac{\phi_{1}}{2}\right)\cos^{2}\left(\frac{\phi_{2}}{2}\right)
    -\cos\left(\frac{\phi_{1}}{2}\right)\sin^{2}\left(\frac{\phi_{2}}{2}\right)
    .
\end{split}
\end{equation*}
To obtain a generic three-qubit exchange-symmetric state, which can be parameterized by a single angle $\phi$ as, $|\phi\rangle_{\Sigma{3}} = \cos{\phi}|0\rangle_{\Sigma{3}} + \sin{\phi}|2\rangle_{\Sigma{3}}$,
we need
\begin{equation}\label{eq:phi1-phi2-eq-N-3-subroutine}
\begin{split}
    \cos{\phi} &= f_{0,0}(\phi_{1},\phi_{2}),\\
    \sin{\phi} &= \frac{\sqrt{3}}{\sqrt{2}}f_{1,1}(\phi_{1},\phi_{2}) = \sqrt{3}f_{0,2}(\phi_{1},\phi_{2}),
\end{split}
\end{equation}
which follows from expanding the three-qubit state, $|2\rangle_{\Sigma{3}}$, in terms of Clebsch-Gordan coefficients as,
\begin{equation*}
    |2\rangle_{\Sigma{3}} 
    = \frac{\sqrt{2}}{\sqrt{3}}|1\rangle_{3}|1\rangle_{\Sigma{2}} + \frac{1}{\sqrt{3}}|0\rangle_{3}|2\rangle_{\Sigma{2}}.
\end{equation*}
Solving Eq.~(\ref{eq:phi1-phi2-eq-N-3-subroutine}) with respect to $\phi_{1}$ and $\phi_{2}$ for a given $\phi$, we first obtain
\begin{equation}\label{eq:tan-phi2-of-theta}
    \tan{\phi_{2}} = \frac{2\sin{\phi}}{\sqrt{3}\cos{\phi} + \sin{\phi}},
\end{equation}
followed by
\begin{equation}\label{eq:tan-phi1-half-of-phi2}
    \tan\left(\frac{\phi_{1}}{2}\right) = \frac{1 -\cos{\phi_{2}}+\sin{\phi_{2}}}{1 +\cos{\phi_{2}}-\sin{\phi_{2}}}.
\end{equation}
Although there are three parameters, namely $\phi$, $\phi_{1}$, and $\phi_{2}$, there is only one degree of freedom in the generated state of the atomic qubits, since they are interdependent.
However, this is sufficient to 
prepare
the most general state of $N=2$ or $N=3$ atomic qubits with even parity, real amplitudes and the exchange symmetry.

For $N = 4$, we still have only one degree of freedom.
To see this, we add an additional qubit as
\begin{equation}\label{eq:zero-ket-4-theta-ket-sigma3}
    |0\rangle_{4}|\phi\rangle_{\Sigma{3}}
    = \cos{(\phi)}|0\rangle_{4}|0\rangle_{\Sigma{3}}
    + \sin{(\phi)}|0\rangle_{4}|2\rangle_{\Sigma{3}}.
\end{equation}
Since swapping the fourth qubit with one of the first three qubits alters the total state, we apply $U_{3}(\phi_{3})$ to make the state symmetric under arbitrary permutation. Since the first three qubits already have the symmetry via the last two layers, $U_{1}(\phi_{1})$ and $U_{2}(\phi_{2})$, a single parameter may be sufficient to entangle the fourth qubit with the rest. 
In the following, we compute this parameter explicitly.
Applying the third layer, we have
\begin{equation}\label{eq:U3-phi3-0ket-4-theta-ket-Sigma3}
\begin{split}
    &U_{3}(\phi_{3})
    |0\rangle_{4}|\phi\rangle_{\Sigma{3}}\\
    &= 
    |0\rangle_{4}
    \cos\left(\frac{\phi_{3}}{2}X_{\Sigma{3}}\right)
    |\phi\rangle_{\Sigma{3}}
    +
    |1\rangle_{4}
    \sin\left(\frac{\phi_{3}}{2}X_{\Sigma{3}}\right)|\phi\rangle_{\Sigma{3}},
\end{split}
\end{equation}
where we use
\begin{equation}\label{eq:Un-into-cos-and-sin}
    U_{n}(\phi)
    = \cos\left(\frac{\phi}{2}X_{\Sigma{n}}\right)
    -i Y_{n+1} 
    \sin\left(\frac{\phi}{2}X_{\Sigma{n}}\right).
\end{equation}
From the parity of cosine and sine functions, 
we can expand
\begin{equation*}
\begin{split}
    \cos\left(\frac{\phi_{3}}{2}X_{\Sigma{3}}\right)
    |\phi\rangle_{\Sigma{3}}
    &\equiv \alpha_{0}|0\rangle_{\Sigma{3}}
    + \alpha_{2}|2\rangle_{\Sigma{3}},\\
    \sin\left(\frac{\phi_{3}}{2}X_{\Sigma{3}}\right)
    |\phi\rangle_{\Sigma{3}}
    &\equiv \beta_{1}|1\rangle_{\Sigma{3}}
    + \beta_{3}|3\rangle_{\Sigma{3}},
\end{split}
\end{equation*}
with the coefficients 
\begin{equation}\label{eq:alpha-even-and-beta-odd-Sigma3}
\begin{split}
    \alpha_{2p}(\phi_{3};\phi)
    &= \sum_{q=0}^{1}{
        {}_{\Sigma{3}}\langle 2p | \cos\left(\frac{\phi_{3}}{2}X_{\Sigma{3}}\right)|2q\rangle_{\Sigma{3}}\langle 2q|\phi\rangle_{\Sigma{3}}
    },\\
    \beta_{2p+1}(\phi_{3};\phi)
    &= \sum_{q=0}^{1}{
        {}_{\Sigma{3}}\langle {2p+1}| \sin\left(\frac{\phi_{3}}{2}X_{\Sigma{3}}\right)|2q\rangle_{\Sigma{3}}\langle 2q|\phi\rangle_{\Sigma{3}}
    }
\end{split}
\end{equation} 
for $p=0,1$.
The matrix elements of $\cos[(\phi_{3}/2)X_{\Sigma{3}}]$ and $\sin[(\phi_{3}/2)X_{\Sigma{3}}]$ can be computed 
via
\begin{equation*} 
    {}_{\Sigma{N}}\langle m | e^{-i\frac{\phi}{2}X_{\Sigma{N}}}|n\rangle_{\Sigma{N}}
    \equiv 
    \sum_{k=0}^{N}
    D_{mn}^{k}
    \cos^{N-k}{\frac{\phi}{2}}
    \sin^{k}{\frac{\phi}{2}},
\end{equation*}
where
\begin{equation*}
    D_{mn}^{k}
    = (-i)^{k}\begin{pmatrix}
        N \\ m
    \end{pmatrix}^{-\frac{1}{2}}
    \begin{pmatrix}
        N \\ n
    \end{pmatrix}^{-\frac{1}{2}}
    \begin{pmatrix}
        N \\ k
    \end{pmatrix}
    \begin{pmatrix}
        N - k \\ m - j_{mnk}
    \end{pmatrix}
    \begin{pmatrix}
        k \\ j_{mnk}
    \end{pmatrix}
\end{equation*}
if $j_{mnk} = (m-n+k)/2$ is an integer in the range,
\begin{equation*}
    \max{\{0,m+k-N\}} \le 
    j_{mnk} \le \min{\{k,m\}},
\end{equation*}
and $D_{mn}^{k} = 0$ otherwise for $0 \le m,n \le N$.
We note that for each given $\phi_{3}$, the angle $\phi$ can be determined to pursue the exchange symmetry of the total state, Eq.~(\ref{eq:U3-phi3-0ket-4-theta-ket-Sigma3}), 
which is now written as
\begin{equation}\label{eq:phi3-theta-ket-expanded}
\begin{split}
    U_{3}(\phi_{3})
    |0\rangle_{4}|\phi\rangle_{\Sigma{3}}
    &= \alpha_{0}|0\rangle_{4}|0\rangle_{\Sigma{3}}
    + \beta_{3}|1\rangle_{4}|3\rangle_{\Sigma{3}}\\
    &+ \alpha_{2}|0\rangle_{4}|2\rangle_{\Sigma{3}}
    + \beta_{1}|1\rangle_{4}|1\rangle_{\Sigma{3}}.
\end{split}
\end{equation}
In order to make this state invariant with respect to any permutation among the four qubits, we first note that the first two terms are already symmetric since
\begin{equation*}
\begin{split}
    |0\rangle_{4}|0\rangle_{\Sigma{3}} &= |0\rangle_{\Sigma{4}},\\
    |1\rangle_{4}|3\rangle_{\Sigma{3}} &= |4\rangle_{\Sigma{4}}.
\end{split}
\end{equation*}
The last two terms in Eq.~(\ref{eq:phi3-theta-ket-expanded}) constitute
\begin{equation}\label{eq:2ket-Sigma4-in-4-and-Sigma3}
    |2\rangle_{\Sigma{4}} = \frac{1}{\sqrt{2}}|0\rangle_{4}|2\rangle_{\Sigma{3}} + \frac{1}{\sqrt{2}}|1\rangle_{4}|1\rangle_{\Sigma{3}}
\end{equation}
if and only if 
\begin{equation}\label{eq:alpha2-equals-beta1}
    \alpha_{2}(\phi_{3};\phi) = \beta_{1}(\phi_{3};\phi).
\end{equation}
From this constraint, we obtain $\phi$ keeping the exchange symmetry after the rotation by $\phi_{3}$.
Computing the coefficients
using Eq.~(\ref{eq:alpha-even-and-beta-odd-Sigma3}) and $|\phi\rangle_{\Sigma{3}} = \cos{\phi}|0\rangle_{\Sigma{3}} + \sin{\phi}|2\rangle_{\Sigma{3}}$, 
the constraint given in Eq.~(\ref{eq:alpha2-equals-beta1}) reads
\begin{equation}\label{eq:alpha2-equals-beta1-explicit}
    (c_{3}+s_{3})(1 - 3c_{3}s_{3}) \sin{\phi} = \sqrt{3} c_{3}s_{3} (c_{3}+s_{3}) \cos{\phi},
\end{equation}
where we define $c_{l} = \cos{(\phi_{l}/2)}$ and $s_{l} = \sin{(\phi_{l}/2)}$.
There are two cases to satisfy this equation: $c_{3}+s_{3} = 0$ or not.
If we set $\phi_{3} = -\pi/2\,\mathrm{mod}\,2\pi$ to have $c_{3} + s_{3} = \sqrt{2}\sin{(\phi_{3}/2 + \pi/4)}$, we get the total state given in Eq.~(\ref{eq:phi3-theta-ket-expanded}) as
\begin{equation*}
    \cos\left(\phi + \frac{\pi}{3}\right)\frac{1}{\sqrt{2}}
    \left[ |0\rangle_{\Sigma{4}} + |4\rangle_{\Sigma{4}} \right]
    - \sin\left(\phi + \frac{\pi}{3}\right)|2\rangle_{\Sigma{4}}.
\end{equation*}
This ansatz has the same probability amplitude for $|0\rangle_{\Sigma{4}}$ and $|4\rangle_{\Sigma{4}}$, suitable for generating the Greenberger-Horne-Zeilinger state, but not optimal for preparing the ground state in the weak-coupling regime where the zero-excitation state is dominant over the rest. Thus, we now seek for the solution of Eq.~(\ref{eq:alpha2-equals-beta1-explicit}) with $c_{3} + s_{3} \neq 0$, for which we have
\begin{equation}\label{eq:phi3-to-theta-for-N4-M}
    \tan{\phi} = \frac{\sqrt{3} \sin{\phi_{3}}}{2 - 3\sin{\phi_{3}}}.
\end{equation}
We note that for $\phi_{3}$ close to zero, the corresponding $\phi$ is also small, so that the total state of the atomic qubits remains close to $|0\rangle_{\Sigma{4}}$. 
Once we determine $\phi$ for each $\phi_{3}$, we sequentially obtain $\phi_{2} = \phi_{2}(\phi_{3})$ from Eq.~(\ref{eq:tan-phi2-of-theta}) and $\phi_{1} = \phi_{1}(\phi_{2})$ via Eq.~(\ref{eq:tan-phi1-half-of-phi2}). Thus, for each given value of $\phi_{3}$, the remaining angles in the previous layers are determined.
We then apply the unitary $U_{n}(\phi_{n})$ with these angles, $\{\phi_{1},\phi_{2},\phi_{3}\}$, to prepare the state Eq.~(\ref{eq:even-num-init-state-generation-subroutine}) for $N=4$ as
\begin{equation}\label{eq:phi3ket-Sigma4-expanded}
    |\phi_{3}\rangle_{\Sigma{4}}^{\mathrm{symm}} 
    \equiv \sum_{p=0}^{2}a_{2p}^{\Sigma 4}(\phi_{3})|2p\rangle_{\Sigma{4}},
\end{equation}
where the three coefficients are determined via Eqs.~(\ref{eq:U3-phi3-0ket-4-theta-ket-Sigma3}), (\ref{eq:2ket-Sigma4-in-4-and-Sigma3}) and (\ref{eq:alpha2-equals-beta1}) as
\begin{equation*}
\begin{split}
    a_{0}^{\Sigma 4}(\phi_{3}) &\equiv \alpha_{0}(\phi_{3};\phi),\\
    a_{2}^{\Sigma 4}(\phi_{3}) &\equiv \sqrt{2}\beta_{1}(\phi_{3};\phi) = \sqrt{2}\alpha_{2}(\phi_{3};\phi),\\
    a_{4}^{\Sigma 4}(\phi_{3}) &\equiv \beta_{3}(\phi_{3};\phi)
\end{split}
\end{equation*}
with $\phi = \phi(\phi_{3})$ given in Eq.~(\ref{eq:phi3-to-theta-for-N4-M}).

In the strong-coupling regime, the atomic state needs to approach

$\frac{1}{\sqrt{2}}|+\rangle_{\Sigma{4}} + \frac{1}{\sqrt{2}}|-\rangle_{\Sigma{4}}$,

where $X_{\Sigma{N}}|\pm\rangle_{\Sigma{N}}={\pm N}|\pm\rangle_{\Sigma{N}}$,
as given in Eq.~(\ref{eq:ground-state-inf-coupling-even-parity}).

To check whether the ansatz can reach this asymptotic limit, we compare this state with Eq.~(\ref{eq:phi3ket-Sigma4-expanded}) and equate
\begin{equation*}
    \alpha_{0}(\phi_{3};\phi) = \beta_{3}(\phi_{3};\phi) = \frac{1}{2\sqrt{2}},
\end{equation*}
whose solution imposes another condition as
\begin{equation*}
    \tan{\phi} = \frac{2 - \sin{\phi_{3}}}{\sqrt{3}\sin{\phi_{3}}}.
\end{equation*}
Combined with Eq.~(\ref{eq:phi3-to-theta-for-N4-M}), it gives
\begin{equation*}
    \sin{\phi_{3}} \xrightarrow{\lambda \gg \lambda_{c}} \frac{1}{2}
\end{equation*}
with the corresponding asymptotic limit of $\phi$ as
\begin{equation*}
    \phi \xrightarrow{\lambda \gg \lambda_{c}} \frac{\pi}{3}\,\mathrm{mod}\,2\pi.
\end{equation*}

For $N=3$ and $N=4$, the exchange symmetry of the generated state could be ensured through the application of $N-1$ layers of gates given in Eq.~(\ref{eq:even-num-init-state-generation-subroutine}), since there is only one necessary condition on the coefficients, e.g., Eq.~(\ref{eq:alpha2-equals-beta1}) for $N=4$. 
From $N=5$, the number of conditions to satisfy becomes more than just one, so that we may need additional layers of gates to compensate the deviation from an exchange-symmetric ansatz.

For general $n \ge 1$, we consider
\begin{equation}\label{eq:phi-n-a-2p-set-ket-Sigma-n-plus-1}
    |{\phi_{n};\{a_{2p}\}}\rangle_{\Sigma,{n+1}}
    \equiv U_{n}(\phi_{n})|0\rangle_{n+1}\sum_{p=0}^{\lfloor{n/2}\rfloor}a_{2p}|2p\rangle_{\Sigma{n}}.
\end{equation}
Expanding $U_{n}$ as Eq.~(\ref{eq:Un-into-cos-and-sin}), we have
\begin{equation}\label{eq:phi-n-a-2p-set-ket-Sigma-n-plus-1-expanded}
\begin{split}
    |{\phi_{n};\{a_{2p}\}}\rangle_{\Sigma,{n+1}}
    &=
    \sum_{q=0}^{\lfloor{n/2}\rfloor}
    \alpha_{2q}|0\rangle_{n+1}|{2q}\rangle_{\Sigma{n}}
    \\
    &+ 
    \sum_{q=1}^{\lfloor{(n+1)/2}\rfloor}
    \beta_{2q-1}|1\rangle_{n+1}|{2q-1}\rangle_{\Sigma{n}},
\end{split}
\end{equation}
where
\begin{equation*}
\begin{split}
    \alpha_{m} \equiv \alpha_{m}(\phi_{n};\{a_{2p}\}) 
    &\equiv \sum_{p=0}^{\lfloor{n/2}\rfloor}
    {}_{\Sigma{n}}\langle{m}|
    \cos\left(\frac{\phi_{n}}{2}X_{\Sigma{n}}\right)|2p\rangle_{\Sigma{n}}a_{2p},\\
    \beta_{m} \equiv \beta_{m}(\phi_{n};\{a_{2p}\}) 
    &\equiv
    \sum_{p=0}^{\lfloor{n/2}\rfloor}
    {}_{\Sigma{n}}\langle{m}|
    \sin\left(\frac{\phi_{n}}{2}X_{\Sigma{n}}\right)|2p\rangle_{\Sigma{n}}a_{2p}
\end{split}
\end{equation*}
for $0 \le m \le n$.
To let the state 
be exchange symmetric, we need
\begin{equation*}
    \alpha_{2q}|0\rangle_{n+1}|{2q}\rangle_{\Sigma{n}}
    + \beta_{2q-1}|1\rangle_{n+1}|{2q-1}\rangle_{\Sigma{n}}
    \propto |2q\rangle_{\Sigma,n+1},
\end{equation*}
where
\begin{equation}\label{eq:2q-ket-Sigma-nplus1-into-two-terms}
\begin{split}
    |2q\rangle_{\Sigma,{n+1}}
    &= \sqrt{\frac{2q}{n+1}}
    |1\rangle_{n+1}|{2q-1}\rangle_{\Sigma{n}}\\
    &+ \sqrt{\frac{n+1-2q}{n+1}}
    |0\rangle_{n+1}|2q\rangle_{\Sigma{n}},
\end{split}
\end{equation}
resulting in
\begin{equation}\label{eq:exch-symm-cond-q}
    \beta_{2q-1}/\alpha_{2q}
    = \sqrt{
        2q/(n+1-2q)
    }
\end{equation}
for $1 \le q \le \lfloor{n/2}\rfloor$.
For a given $\phi_{n}$, this translates into a system of linear equations with respect to $\{a_{2p}\}$ as
\begin{equation*}
    0 
    = \sum_{p=0}^{\lfloor{n/2}\rfloor}
    \left[
    \sqrt{n+1-2q}S_{2q-1,2p}(\phi_{n}) - \sqrt{2q}C_{2q,2p}(\phi_{n})
    \right]a_{2p},
\end{equation*}
where $C_{m,2p}(\phi) + i S_{m,2p}(\phi) \equiv {}_{\Sigma{n}}\langle{m}|\exp\left(i{\phi}X_{\Sigma{n}}/2\right)|2p\rangle_{\Sigma{n}}$.
The solution belongs to the null space of the matrix representing the linear system. 
For $n=4$ as an example, it is a $2 \times 3$ real matrix which can be written as
\begin{equation*}
\begin{bmatrix}
    \sqrt{3}S_{10} - \sqrt{2}C_{20} & \sqrt{3}S_{12} - \sqrt{2}C_{22} & \sqrt{3}S_{14} - \sqrt{2}C_{24}\\
    \sqrt{1}S_{30} - \sqrt{4}C_{40} & \sqrt{1}S_{32} - \sqrt{4}C_{42} & \sqrt{1}S_{34} - \sqrt{4}C_{44}
\end{bmatrix}
\end{equation*}
for a given $\phi_{4}$. Since there are two constraints (rows) on the three normalized coefficients, namely $a_{0}$, $a_{2}$, and $a_{4}$, the solution can be determined up to a global sign as long as all the rows are linearly independent. For instance, when $\phi_{4} = 0$, we have $C_{22} = C_{44} = 1$ with all the other terms vanishing so that the null space of the matrix is spanned by a unit vector, $(1,0,0)$, identifying $(a_{0},a_{1},a_{2}) = (\pm 1, 0, 0)$.
Denoting the solution as $\{a_{2p}^{0}(\phi_{n})\}$ for a given $\phi_{n}$,
with the global sign fixed via $a_{0}^{0}(0) = 1$, the state given in Eq.~(\ref{eq:phi-n-a-2p-set-ket-Sigma-n-plus-1}) becomes symmetric by setting $a_{2p} = a_{2p}^{0}(\phi_{n})$ as
\begin{equation*}
\begin{split}
    |{\phi_{n}}\rangle_{\Sigma,{n+1}}^{\mathrm{symm}}
    &\equiv |{\phi_{n};\{a_{2p}^{0}(\phi_{n})\}}\rangle_{\Sigma,{n+1}}\\
    &= U_{n}(\phi_{n})|0\rangle_{n+1}|\phi_{n}\rangle_{\Sigma{n}}^{0},
\end{split}
\end{equation*}
where 
\begin{equation*}
    |\phi_{n}\rangle_{\Sigma{n}}^{0} \equiv \sum_{p=0}^{\lfloor{n/2}\rfloor}a_{2p}^{0}(\phi_{n})|2p\rangle_{\Sigma{n}}.
\end{equation*} 
The coefficients 
of the symmetric state can be computed via Eqs.~(\ref{eq:phi-n-a-2p-set-ket-Sigma-n-plus-1-expanded}) and (\ref{eq:2q-ket-Sigma-nplus1-into-two-terms}) as
\begin{equation*}
\begin{split}
    a_{n+1,2q}(\phi_{n}) &\equiv {}_{\Sigma,n+1}\langle{2q}|{\phi_{n}}\rangle_{\Sigma,{n+1}}^{\mathrm{symm}}\\
    &= \alpha_{2q}\sqrt{\frac{n+1-2q}{n+1}} + \beta_{2q-1}\sqrt{\frac{2q}{n+1}},
\end{split}
\end{equation*}
where $0 \le q \le \lfloor{(n+1)/2}\rfloor$.

In short, in order for the state given in Eq.~(\ref{eq:phi-n-a-2p-set-ket-Sigma-n-plus-1}) to be exchange symmetric, the initial amplitudes should be prepared as $a_{2p} = a_{2p}^{0}(\phi_{n})$ for a given $\phi_{n}$.
For $N \le 4$, it is possible to prepare $|\phi_{n}\rangle_{\Sigma{n}}^{0}$ for any $n \le N-1$ so that $|\phi_{n}\rangle_{\Sigma,n+1}^{\mathrm{symm}}$ can be realized, since the number of normalized coefficients in $|\phi_{n}\rangle_{\Sigma{n}}^{0}$ is $1 + \lfloor{n/2}\rfloor \le 2$, which can be parameterized by up to one angle.
For $N > 4$, there are more than two coefficients to prepare $|\phi_{n}\rangle_{\Sigma{n}}^{0}$, in which case we need more than one degree of freedom. 
In fact, for a given $\phi_{n}$, there is in general no $\phi_{n-1}$ satisfying $a_{n,2p}(\phi_{n-1}) = a_{2p}^{0}(\phi_{n})$ for all $p$ if $n \ge 4$. 
However, for $n\ge 2$, we observe an optimal value of $\phi_{n}$ such that the difference between the coefficients are sufficiently small. We determine $\phi_{n-1}$ by minimizing the infidelity $1 - \prescript{0}{\Sigma{n}}{\langle \phi_{n}|}\phi_{n-1}\rangle_{\Sigma{n}}^{\mathrm{symm}}$ for each $\phi_{n}$.

\bibliography{ref}
\end{document}